\documentclass[preprint,12pt]{elsarticle}

\usepackage{epsfig}
\usepackage{graphicx}
\usepackage{float}
\usepackage{url}

\usepackage[labelsep=period]{caption}
\usepackage{subcaption}
\usepackage{amsmath}
\usepackage{microtype}
\usepackage{multirow}
\usepackage{amsmath,amsfonts,amssymb,amsthm}
\usepackage{mathtools}
\usepackage{commath}
\usepackage[sc,osf]{mathpazo}
\usepackage[super]{nth}
\usepackage{verbatim} 
\usepackage{pgf}
\usepackage{pgfplots, pgfplotstable}

\DeclareMathAlphabet{\mathpzc}{OT1}{pzc}{m}{it}
\usepackage{pgffor}
\usepackage{parskip}

\pgfplotsset{compat=1.18}

\RequirePackage{import}

\newcommand{\DT}{{I}}                  

\newcommand{\SDvect}[1]{%
            \underline{\boldsymbol{
                {#1}
            }}
}

\newcommand{\SDtens}[1]{%
            \underline{\underline{\boldsymbol{
                {#1}
            }}}
}
 
\newcommand{\normi}[1]{\left\|{#1}\right\|}

\usepackage{algorithm}
\usepackage{algpseudocode}
\usepackage{hyperref}

\journal{Mechanical Systems and Signal Processing}

\begin{document}

\begin{frontmatter}




\title{Physics-Informed Single Atom Convolutional Matching Pursuit: Guided-Waves Wavenumbers and Propagation Distance Estimation for Damage Localization in Structural Health Monitoring}

\author[label1]{Sebastian Rodriguez \corref{mycorrespondingauthor}}
\cortext[mycorrespondingauthor]{Corresponding author}
\ead{sebastian.rodriguez_iturra@ensam.eu}
\author[label1]{Borja Ferrandiz}
\author[label1,label2]{Francisco Chinesta}
\author[label1]{Nazih Mechbal}
\author[label1]{Marc Rébillat}

\address[label1]{{PIMM, Arts et Métiers ParisTech, CNRS, CNAM},
            {151 Boulevard de l'Hôpital}, 
            {Paris},
            {75013}, 
        		{France}}

\address[label2]{{CNRS@CREATE LTD},
            {1 Create Way, \#08-01 CREATE Tower}, 
            {Singapore},
            {138602}, 
            {Singapore}}



\begin{abstract}

Structural Health Monitoring (SHM) aims at the real-time monitoring of the integrity of engineering structures, with Guided-waves (GWs) providing high sensitivity to damage presence and to ageing effects for thin-walled components. In conventional GW-based SHM, a bonded piezoelectric transducer (PZT) emits a short tone burst that produces an Initial Wave Packet (IWP) propagating through the structure. As this packet interacts with boundaries and potential damages, additional scattered wave packets are produced. A major limitation of such approaches lies in the simultaneous excitation of multiple dispersive GW modes by a single PZT, which significantly complicates signal interpretation and damage monitoring. In this context, this work proposes the Physics-Informed Single Atom Convolutional Matching Pursuit (PISACMP) method, a signal decomposition method grounded in the physical principles governing wave propagation. In contrast with purely data-driven or numerically intensive techniques, the proposed approach embeds strong physical constraints into a low-dimensional and computationally efficient signal representation. This formulation enables the direct identification of key physically meaningful features, including modal wavenumber functions and propagation distances between actuator, damage and sensors. These extracted features, especially source-damage-sensor distances, allows to subsequently perform damage location using well established Elliptical Localization techniques. The principal novelty of this study lies in integrating wave propagation physics into a compact signal decomposition framework and developing an interpretable damage localization methodology for GW-SHM applications.

\end{abstract}




\end{frontmatter}


\section{Introduction}
\label{sec:Introduction}

In industrial applications, ensuring the structural integrity of critical components is a major engineering challenge. The automatic monitoring of damage is essential to prevent catastrophic failures and reduce maintenance costs. In this context, damage refers to changes in material properties and/or structural geometry that adversely affect the current or future performance of a structure. The systematic process of detecting, localizing, and characterizing such changes is known as Structural Health Monitoring (SHM) \citep{worden_fundamental_2007, sohn_review_2002,guo2023guided,di2024damage,galanopoulos2025shm}.

Among the various SHM techniques developed for thin metallic and composite structures, approaches based on ultrasonic Guided Waves (GW) generated and sensed by bonded piezoelectric transducers (PZT) have proven particularly effective \citep{su_guided_2006, su_identification_2009, mitra_guided_2016, qing_piezoelectric_2019}. In a typical GW-based SHM configuration, a PZT actuator bonded to the structure emits a tone burst centered around a selected frequency. This excitation generates an initial wave packet (IWP) that propagates through the structure and interacts with boundaries, geometric discontinuities, and potential damages. Each interaction produces additional scattered wave packets, resulting in measured signals that contain overlapping and dispersed contributions from multiple propagation paths. Detecting and isolating damage-related echoes within such complex signals is therefore a central challenge. In this framework, the ability to decompose measured signals into physically interpretable wave packets that can be directly associated with structural features is of paramount importance.

Several signal processing strategies have been proposed to address this issue. These approaches can be broadly classified into dictionary-based and non-dictionary-based methods. \textit{Dictionary-based} techniques include, for instance, the Matching Pursuit Method (MPM) \citep{mallat1993matching}, which approximates a signal as a linear combination of predefined atoms selected from an over-complete dictionary. In guided waves applications, such dictionaries must be constructed a priori to account for delayed, attenuated, and dispersed versions of both A0 and S0 modes \citep{xu2009lamb, raghavan2007guided, chakraborty2009damage, lu2008numerical, mu2021ultrasound}. However, accurately precomputing all possible modal contributions for continuous thin structures is impractical, limiting the method’s adaptability and physical consistency. \textit{Non-dictionary-based} approaches—such as Empirical Mode Decomposition (EMD), Variational Mode Decomposition (VMD), Proper Orthogonal Decomposition (POD), and Hilbert–Huang Transform methods—have also been applied to guided wave analysis \citep{cuomo2023damage, jiang2023quantitative}. Recent work introduced the Rank Reduction AutoEncoder (RRAE), a purely data-driven architecture that automatically extracts features from measured signals to maximize damage detection performance \cite{rodriguez2026damage}. While these techniques can effectively separate multi-component signals, their mathematical formulations are often weakly connected to the physics of wave propagation, limiting the physical interpretability of the extracted components. 


To overcome these limitations, this work introduces a decomposition framework grounded in the mathematical description of guided wave propagation in thin plates. This propagation is modeled in the frequency domain through a wavenumber function that depends on frequency and propagation distance. On one hand, the wavenumber function captures dispersive effects, namely the propagation induced distortion of the Initial Wave Packet (IWP). On the other hand, the propagation distance accounts for the extent to which the signal travels through the solid medium. Therefore, the proposed approach models measured signals as delayed and dispersed impulse responses of the external excitation, thereby directly reflecting the fundamental physics of guided waves propagation. Its primary objective is to estimate not only the modal wavenumber functions of the structure, but also the propagation distances traveled by the different wave packets from the original source to the sensing locations. By embedding physical constraints into a compact and low-dimensional signal representation, the method enables the simultaneous extraction of these physically meaningful parameters while preserving computational efficiency and robustness.

The proposed decomposition, referred to as the Physics-Informed Single Atom Convolutional Matching Pursuit (PISACMP) method, can be viewed as a physics-driven extension of the Single Atom Convolutional Matching Pursuit framework introduced in \cite{rodriguez2025single}. Beyond modal characterization, the method provides estimates of propagation distances between actuator, damage, and sensors. As an illustration of their usefulness for SHM purposes, in this work, these estimated distances are subsequently employed to perform efficient damage location using well-established Elliptical Localization techniques, enabling accurate and physically consistent detection of damages in thin plate structures.

The remainder of this paper is organized as follows. Section~\ref{sec:ref_prob} presents the reference problem addressed in this study, namely the propagation of guided waves in thin plates. Section~\ref{sec:PIMPM} introduces the proposed physics-informed numerical methodology for estimating both modal wavenumber functions and the propagation distances traveled by wave packets from the source, based solely on simulated signal data. Next Section~\ref{sec:PISACMP_dam_loc} presents the application of PISACMP for efficient and interpretable damage localization. Section~\ref{sec:numerical_results} then evaluates the performance of the approach through a series of numerical investigations, including the approximation of both simple and complex signals and its application to damage localization. Finally, Section~\ref{sec:concl_pers} summarizes the main findings and outlines perspectives for future research.

\section{Reference problem: guided-waves propagation in an isotropic homogeneous medium}\label{sec:ref_prob}

\begin{figure}[H]
\centering
\includegraphics[width=0.65\textwidth]{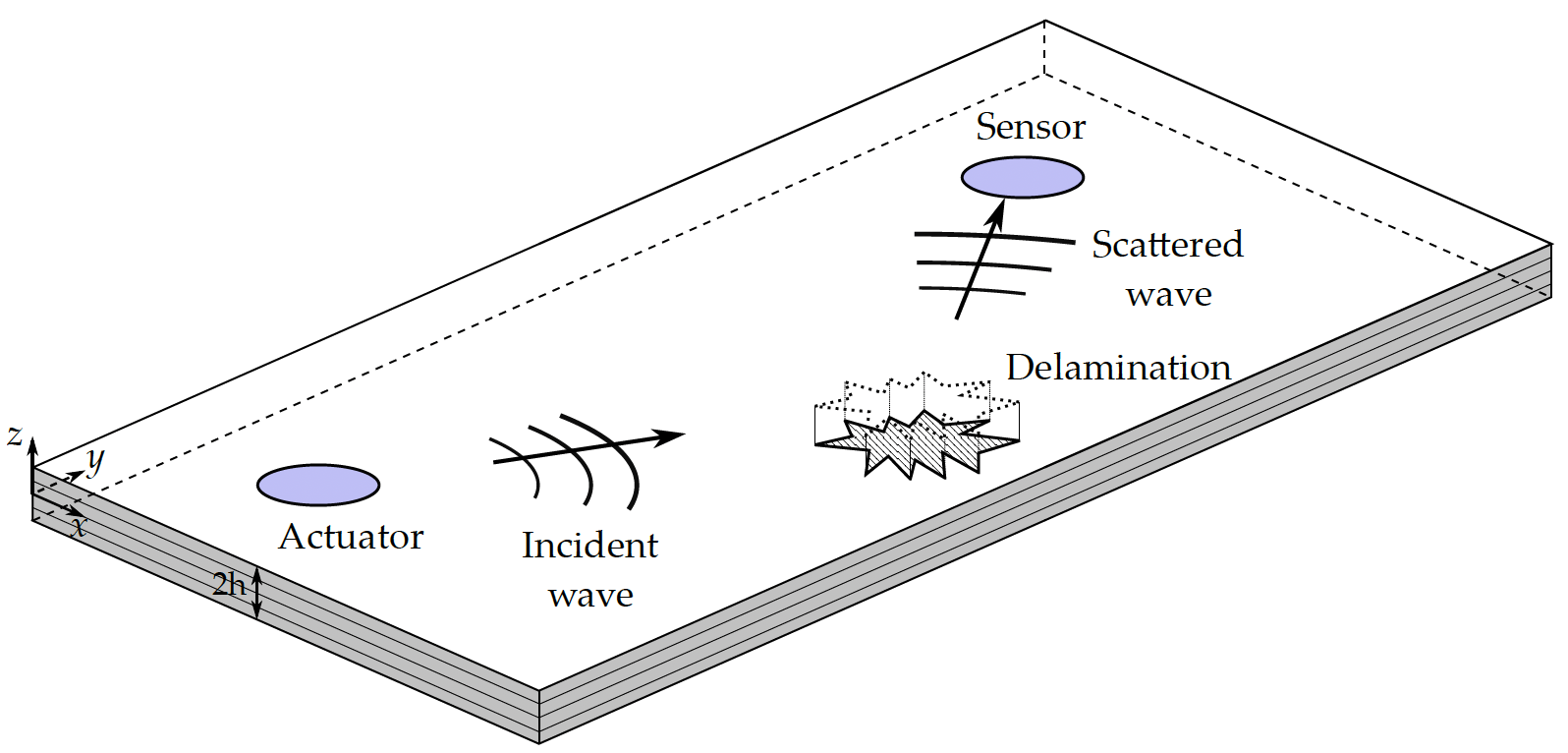}
\caption{Wave propagation in a plate.}
\label{fig:ref_prob_shm}
\end{figure}

The reference problem considered in the following consists on the processing of guided-waves ultrasonic signals that propagates over thin plates made out of isotropic homogeneous materials as is shown in Fig.~\ref{fig:ref_prob_shm}. Such a plate merely corresponds to a composite plate commonly used in aeronautic structures.
In the low frequency range of such structure, only two wave modes exist, namely the $A_0$ and $S_0$ modes, and given an input frequency $f$, the corresponding wavenumbers $k_{S_0}(f)$ and $k_{A_0}(f)$ can be computed as \citep{su_guided_2006, su_identification_2009, mitra_guided_2016, qing_piezoelectric_2019}:
\begin{equation}\label{eq:wav_non_disp}
    k_{S_0}(f) = 2\pi f \sqrt{\frac{\rho}{Q}}
\end{equation}
\begin{equation}\label{eq:wav_disp}
    k_{A_0}(f) = \frac{2\pi f}{\sqrt{2}}
    \sqrt{
    \left(\frac{\rho}{Q}+\frac{\rho}{G\xi}\right)
    + \sqrt{
    \left(\frac{\rho}{Q} - \frac{\rho }{G\xi}\right)^2 + 
    \frac{1}{\pi f^2} \left(\rho + \frac{\rho }{I Q} \right)
    }
    }
\end{equation}
with: $G = \frac{E}{2(1+\nu)}$, $Q = \frac{E}{(1-\nu^2)}$, $\xi=\pi^2/12$ and $I = h^3/12$. $E,\nu,h,\rho$ denotes the homogenized Young modulus, Poisson modulus, thickness of the plate and density of the material respectively. 

\begin{figure}[H]
\centering
\includegraphics[width=0.9\textwidth]{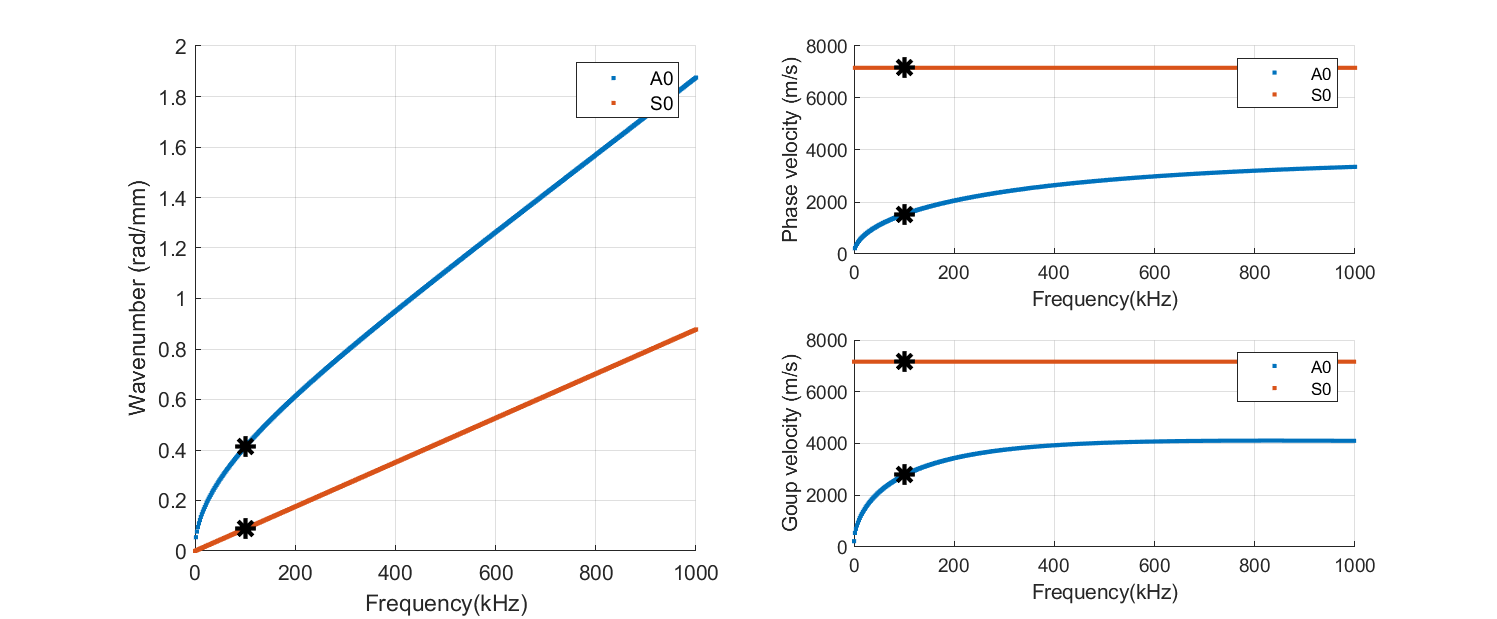}
\caption{Dispersion curves for the simulated structure under study. [Left] Wavenumber $k$ versus frequency. [Top right] Phase velocity versus frequency. [Bottom right] Group velocity versus frequency. The black symbols denotes $A_0$ and $S_0$ waves properties at the selected input frequency of $f_0=100$~[kHz].}
\label{fig:wavenumber_func}
\end{figure}

For a plate with thickness $h = 2$~[mm], Young's modulus $E = 70$~[GPa], Poisson's ratio $\nu = 0.3$, and density $\rho = 1500$~[kg/m$^3$], the wavenumber functions $k_{S_0}(f)$ and $k_{A_0}(f)$ are shown in Fig.~\ref{fig:wavenumber_func}. On this figure, the two dispersion branches corresponding to the $A_0$ and $S_0$ modes can be seen as well as the phase and group velocities. The phase velocity of the $S_0$ mode is not changing with frequency, meaning that the $S_0$ mode does not endure dispersion, whereas the one of the $A_0$ mode is enduring dispersion. From Fig.~\ref{fig:wavenumber_func}, one can conclude that a given signal that is measured by sensors in the structure in theory could be approximated by a sum of signals that almost don't undergo dispersion and ones that undergo dispersion.

Mathematically speaking, by considering a given input signal $x(t)$ (or equivalently its Fourier transform $\hat{x}(\omega)$), the propagated signal at a distance $d$ can be computed as:
\begin{equation}\label{eq:phy_wave_prop_analytical}
 s_{d}(t) = \mathcal{F}^{-1} \left[\displaystyle \sum_{n}^{S_{0}, A_{0}}  \hat{x}(\omega) e^{-j k_n(\omega) d} \right]
\end{equation}
where $\mathcal{F}^{-1}$ denotes the inverse Fourier transform and $j = \sqrt{-1}$.

\begin{figure}[H]
\centering
    \includegraphics[width=\textwidth]{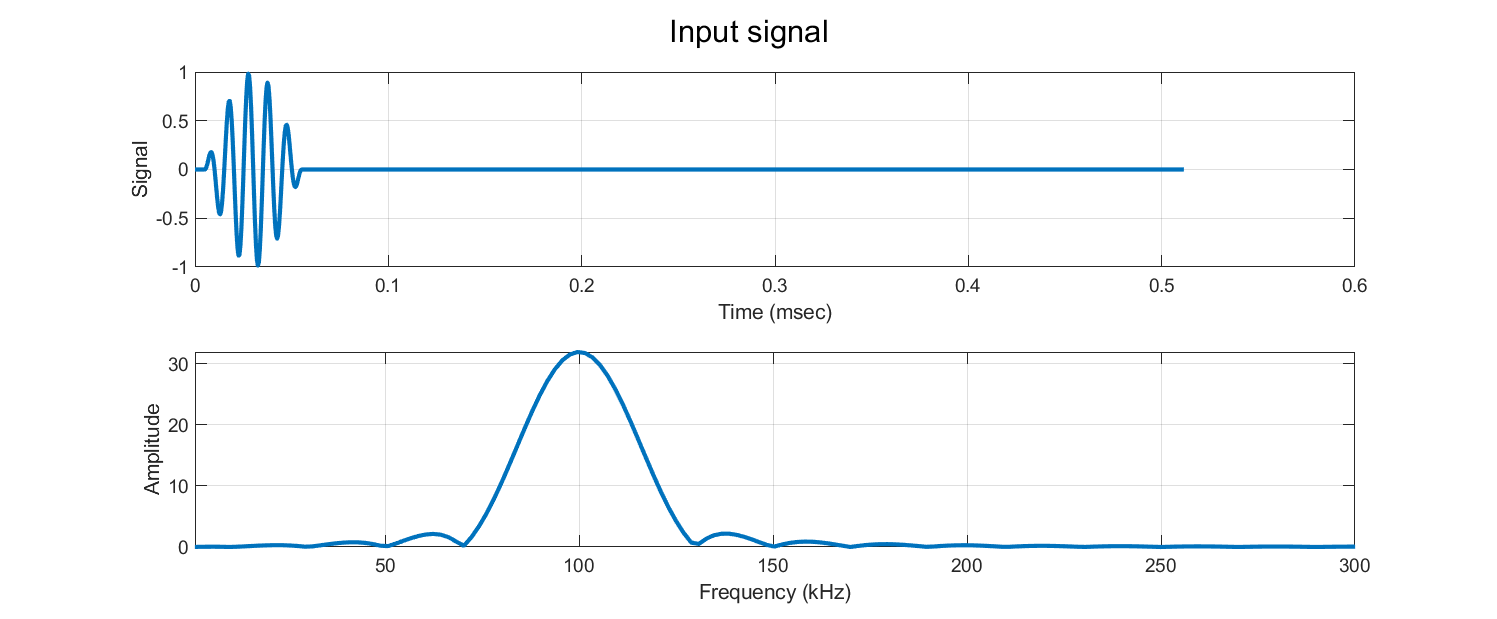}
    \caption{Typical input signal used for LW SHM purposes. Here the central frequency $f_0$ is $100$~[kHz] and the burst is composed of $5$ cycles with a half-sinusoidal window.}
    \label{fig:input}
\end{figure}

\begin{figure}[H]
\centering
\includegraphics[width=0.65\textwidth]{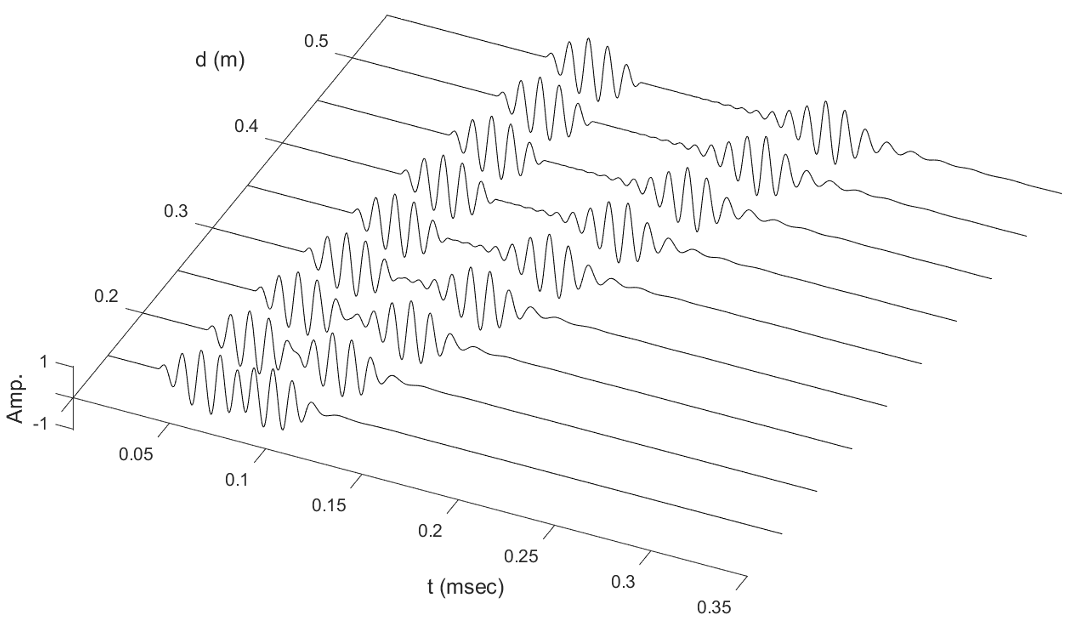}
\caption{Simulated signals for various propagating distances.}
\label{fig:dispersion_of_waves}
\end{figure}

To illustrate this, let us consider the input signal $x(t)$ shown in Fig.~\ref{fig:input}, which consists on a burst signal with a central frequency $f_0=100$~[kHz] and with $5$~cycles bursts.
In this case, signals have been computed for distances $d$ ranging from $15$~[cm] to $55$~[cm] by step of $5$~[cm]. The resulting signals corresponding to these computations are shown in Fig.~\ref{fig:dispersion_of_waves}. 
As can be seen from this figure, the signals are made up of 2 wave packets: the first one, the fastest, corresponds to $S_0$, and the second one, the slowest, to $A_0$. From Fig.~\ref{fig:wavenumber_func} , at the selected input frequency of $f_0=100$~[kHz], it can clearly be seen that the structure is excited in a frequency region where the $S_0$ wave packet does not suffer from dispersion, whereas the $A_0$ one is distorted during propagation because of dispersion.

An important aspect to emphasize is that dispersion phenomena—namely, the progressive distortion of the signal as it propagates through the solid medium—are entirely encoded in the wavenumber functions. This compact yet powerful formulation provides a physically rigorous description of wave propagation. Moreover, the formulation can be interpreted as a Matching Pursuit–type decomposition, in which a single atom $x(t)$ is employed to represent the measured signal at the sensor level. 

However, its direct application requires prior knowledge of the wavenumber functions, which is generally unavailable in complex or realistic structural configurations. To address these limitations, the following sections introduce a physics-informed signal decomposition framework that enables the numerical estimation of the underlying wavenumber functions directly from measured signals. By incorporating the physics of guided wave propagation into the decomposition process, the proposed approach yields a physically meaningful representation of the measured data. This enhanced formulation provides a solid foundation for the subsequent Sections for the application of a robust Guided-waves based Structural Health Monitoring technique.

\section{Physics-Informed Single Atom Convolutional Matching Pursuit (PISACMP)}\label{sec:PIMPM}

Inspired by expression \eqref{eq:phy_wave_prop_analytical}, here we propose the following approximation to reconstruct any signal measured by a sensor given an excitation $\hat{x}(\omega)$ as follows:
\begin{equation}\label{eq:phy_wave_prop}
s(t) \approx \mathcal{F}^{-1} \left[\displaystyle  \sum_{i=1}^{N} \alpha_i(d_i) \hat{x}(\omega) e^{-j k_i(\omega) d_i} \right]
\end{equation}
with $k_i(\omega) \in \mathbb{R}$. Attenuation is modelled by considering the amplitude $\alpha_i(d_i)$.

From the above, one can also write:
\begin{equation}
\hat{s}(\omega) \approx  \sum_{i=1}^{N} \alpha_i(d_i) \hat{x}(\omega) e^{-j k_i(\omega) d_i}  
\end{equation}
with $\hat{s}(\omega) = \mathcal{F}(s(t))$.

Here, $N$ denotes the number of terms of the decomposition. The wavenumber functions $k_{i}(\omega)$ and the distances $d_i$ are computed numerically on-the-fly by minimizing the following error:
\begin{equation}\label{eq:min_prob_freq}
\lbrace   k_i(\omega) , d_i, \alpha_i   \rbrace_{i=1}^{N} = \\ \underset{\lbrace k_i(\omega) , d_i, \alpha_i \rbrace_{i=1}^{N}}{\text{arg min}}
    \normi{ \hat{s}(\omega) -  \sum_{i=1}^{N} \alpha_i \hat{x}(\omega) e^{-j k_i(\omega) d_i} }_{\DT_{\omega}}^{2}
\end{equation}
with the norm in the frequency domain defined as:
\begin{equation}
\normi{\bullet}_{\DT_{\omega}}^2 = \int_{\DT_{\omega}} (\bullet)^* (\bullet) d\omega   
\end{equation}
where $\DT_{\omega}$ denotes the frequency interval domain and $(\bullet)^*$ the complex conjugate of $\bullet$.

Let us notice that solving problem \eqref{eq:min_prob_freq} involves the resolution of a complex nonlinear problem, since the wavenumber function is inside an exponential. In addition to this, $N$ terms should be determined, drastically increasing the resolution complexity. In this context, here we chose to solve each term of \eqref{eq:phy_wave_prop} incrementally one after the other in a Greedy process. This strategy of calculation is given in the following sections.

\textbf{Remark 1}: The above decomposition can be directly related to the Single Atom Matching Pursuit Method (SAMPM) and its convolutional extension (SACMPM) \citep{rodriguez2025single}. Indeed, all the signals computed considering $k(\omega) = c \omega$ with $c \in \mathbb{R}$ exactly corresponds to SAMPM, in where the terms built only correspond to the translation and amplitude modification of the input signal $x(t)$, while the one considering $k(\omega)$ different from linear corresponds to a generalization of the SACMPM method with no physical constraints put on $k(\omega)$ when converting the generated terms into the temporal domain.

\textbf{Remark 2}: Although the method proposed herein is established for wave propagation in isotropic materials, it remains applicable to composite materials provided that an equivalent homogenized representation can be adopted. For structures exhibiting strong anisotropy, for which homogenization techniques are no longer suitable, the PISACMP method can still be implemented. In this context, however, the wavenumber functions and the associated distance measures may be significantly altered by the directional dependence of the material properties. The treatment of such effects is beyond the scope of the present work and will be investigated in future studies.

\subsection{Initialization of the algorithm for the determination of the "$i$-th" term}

The algorithm begins by initializing the wavenumber function. For simplicity, the non-dispersive wavenumber given in Eq. \eqref{eq:wav_non_disp} is selected as the initial estimate. Thus, the initial wavenumber is defined as:
\begin{equation}\label{eq:init_k}
k_i^{[0]}(\omega) = k_{S_0}(\omega) = \omega \sqrt{\frac{\rho}{Q}}
\end{equation}
with $\omega = 2 \pi f$ and $Q = \frac{E}{(1-\nu^2)}$. Here, $\rho$, $E$ and $\nu$ denote the homogenized density, Young’s modulus, and Poisson’s ratio, respectively. These quantities should be determined for each respective problem. However, when they are not available, one can initialize by simply considering:
\begin{equation}
k_i^{[0]}(\omega) = \omega
\end{equation}

\subsection{Determination of amplitude $\alpha_i$ and distance $d_i$ at iteration $\ell$}\label{sec:amp_distance_PIMPM}

Since the problem is nonlinear, we employ a fixed-point iterative strategy for its resolution, where $\ell$ denotes the iteration index.

For the determination of the amplitude $\alpha_i$ we minimize the following error, given $k_i^{[\ell -1]}(\omega)$ known from previous calculation:
\begin{equation}
\alpha_i^{[\ell]}(d)  = \\ \underset{ \alpha_i^{[\ell]}(d) }{\text{arg min}}
    \normi{ \hat{s}^{[i-1]}_{r}(\omega) -   \alpha_i^{[\ell]}(d) \hat{x}(\omega) e^{-j k_i^{[\ell -1]}(\omega) d} }_{\DT_{\omega}}^{2}
\end{equation}

By minimizing the above error with respect to the amplitude, one obtains:
\begin{equation}\label{eq:amp_d}
\alpha_i^{[\ell]}(d) = \mathcal{R} \left( \frac{ \int_{\DT_{\omega}} \left[ \hat{x}(\omega) e^{-j k_i^{[\ell -1]}(\omega) d} \right]^* \hat{s}^{[i-1]}_{r}(\omega)  d\omega  }{  \int_{\DT_{\omega}} \left[ \hat{x}(\omega) e^{-j k_i^{[\ell -1]}(\omega) d} \right]^* \left[ \hat{x}(\omega) e^{-j k_i^{[\ell -1]}(\omega) d} \right] d\omega } \right)
\end{equation}
with $\mathcal{R}(\bullet)$ the real part of $\bullet$.

Let's notice that the amplitude that minimize the error depends on a given distance $d$.

Now, to obtain the optimal amplitude and distance that minimize the overall error, one should minimize the following:

\begin{equation}\label{eq:z_d}
z_{i}(d) = 
\normi{ \hat{s}^{[i-1]}_{r}(\omega) -  \alpha_i^{[\ell]}(d) \hat{x}(\omega) e^{-j k_i^{[\ell -1]}(\omega) d}   }_{\DT_{\omega}}^{2}
\end{equation}

By replacing \eqref{eq:amp_d} into \eqref{eq:z_d} and simplifying the terms, the previous equation can be simplified in:
\begin{equation}
z_{i}(d) = \int_{\DT_{\omega}} \hat{s}^{[i-1]}_{r}(\omega)^* \hat{s}^{[i-1]}_{r}(\omega) d\omega - \frac{\left[ \int_{\DT_{\omega}} \left[ \hat{x}(\omega) e^{-j k_i^{[\ell -1]}(\omega) d} \right]^* \hat{s}^{[i-1]}_{r}(\omega)  d\omega \right]^2}{   \int_{\DT_{\omega}} \left[ \hat{x}(\omega) e^{-j k_i^{[\ell -1]}(\omega) d} \right]^* \left[ \hat{x}(\omega) e^{-j k_i^{[\ell -1]}(\omega) d}  \right] d\omega }
\end{equation}

Therefore, $d_i^{[\ell]}$ is simply calculated as the positive scalar number that minimize $z_{i}(d)$.
\begin{equation}
d_i^{[\ell]} = \underset{ d }{\text{arg min}}( z_{i}(d) )    
\end{equation}
which in turn, due to \eqref{eq:amp_d} automatically defines the best amplitude $\alpha_i^{[\ell]}$.

\subsection{Determination of $k_i(\omega)$ at iteration $\ell$}\label{sec:wavenumber_func}

Given the distance and amplitude $d_i$ and $\alpha_i$ respectively at iteration $\ell$, we compute here an enrichment of the wavenumber function at iteration $\ell$. Since we are dealing with an exponential function containing the wavenumber, its calculation should be performed by linearisation. In this sense, let's considers its first order Taylor approximation at iteration $\ell$ as:
\begin{equation}
e^{-j k_{i}^{[\ell]}(\omega) d_i} \approx e^{-j k_{i}^{[\ell-1]}(\omega) d_i} \left( 1 - j d_i \Delta k_{i}^{[\ell]}(\omega)  \right)
\end{equation}
such as:

$\forall i$,
\begin{equation*}
k_{i}^{[\ell]}(\omega) \leftarrow k_{i}^{[\ell-1]}(\omega) + \Delta k_{i}^{[\ell]}(\omega)
\end{equation*}
This is, at each iteration $\ell$ the wavefunction is updated.

In this sense, one can reformulate problem \eqref{eq:min_prob_freq} in order to compute the corrector $\Delta k_{i}^{[\ell]}(\omega)$ as follows:
\begin{equation}\label{eq:orig_problem}
\Delta k_{i}^{[\ell]}(\omega) = \underset{\Delta k_{i}^{[\ell]}(\omega)}{\text{arg min}}
    \normi{ \hat{s}^{[i-1]}_{r}(\omega) - \alpha_i \hat{x}(\omega) e^{-j k_{i}^{[\ell-1]}(\omega) d_i } \left( - j d_i \Delta k_{i}^{[\ell]}(\omega)  \right) }_{\DT_{\omega}}^{2}
\end{equation}
with the residual function given by:
\begin{equation}
\hat{s}^{[i-1]}_{r}(\omega) = \hat{s}(\omega) - \sum_{m=1}^{i-1} \alpha_m \hat{x}(\omega) e^{-j k_{m}(\omega) d_m} 
\end{equation}


Problem \eqref{eq:orig_problem}, can be rewritten as follows:

$\forall i \in [1,N]$,
\begin{equation}\label{eq:Deltak_prob}
\Delta k_{i}^{[\ell]}(\omega) = \underset{\Delta k_{i}^{[\ell]}(\omega)}{\text{arg min}}
    \normi{ \hat{s}_{r}^{[i-1]}(\omega) - a^{[\ell-1]}_{i}(\omega) \left( - j d_i \Delta k_{i}^{[\ell]}(\omega) \right)  }_{\DT_{\omega}}^{2}
\end{equation}
with:
\begin{equation}
a^{[\ell-1]}_{i}(\omega) = \alpha_i \hat{x}(\omega) e^{-j k_{i}^{[\ell-1]}(\omega) d_i }
\end{equation}
To solve $\Delta k_i^{[l]}(\omega)$ while ensuring a smooth and regular solution, we approximate it using Chebyshev polynomials of the second kind (due to their numerical efficiency) as shape functions:
\begin{equation}\label{eq:Chev_approx}
    \Delta k^{[\ell]}_{i}(\omega) = \sum_{h=1}^{N_c} \Delta \beta^{[\ell]}_{i,h} N_{h}(\omega)
\end{equation}
with $N_{h}(\omega)$ the Chebyshev polynomials defined in the frequency domain.

By minimizing \eqref{eq:Deltak_prob}, one obtains:

$\forall \ \delta \Delta k_{i}^{[\ell]}(\omega)$,
\begin{equation}\label{eq:min_dev}
\begin{split}
&\int_{\DT_{\omega}} \left[  a^{[\ell-1]}_{i}(\omega) \left( - j d_i \delta \Delta k_{i}^{[\ell]}(\omega)  \right) \right]^* \\ &\left[  \hat{s}_{r}^{[i-1]} (\omega) - a^{[\ell-1]}_{i}(\omega) \left( - j d_i \Delta k_{i}^{[\ell]}(\omega)  \right) \right] d\omega = 0 
\end{split}
\end{equation}

By introducing \eqref{eq:Chev_approx} in \eqref{eq:min_dev}, the above formulation can be reduced on the following discretized system of equations:
\begin{equation}\label{eq:Matrix_eq_k}
\SDtens{M}^{[\ell]}_i \Delta \SDvect{\beta}^{[\ell]}_i = \SDvect{f}^{[\ell]}_i
\end{equation}
in where:
\begin{equation}
\SDtens{M}^{[\ell]}_i = \int_{\DT_{\omega}} \left[ a^{[\ell-1]}_{i}(\omega) \left( - j d_i \sum_{h=1}^{N_c}  N_{h}(\omega) \right)
\right]^* \left[ a^{[\ell-1]}_{i}(\omega) \left( - j d_i \sum_{h=1}^{N_c}  N_{h}(\omega) \right) \right] d\omega
\end{equation}
and
\begin{equation}
\SDvect{f}^{[\ell]}_i = \int_{\DT_{\omega}} \left[ a^{[\ell-1]}_{i}(\omega) \left( - j d_i \sum_{h=1}^{N_c}  N_{h}(\omega) \right)
\right]^* \hat{s}_{r}^{[i-1]}(\omega) \  d\omega 
\end{equation}

Solving \eqref{eq:Matrix_eq_k} finally yields the corrective wavenumber through \eqref{eq:Chev_approx}.

\textbf{Remark}: In general, the use of $N_c = 6$ Chebyshev shape functions is recommended to ensure sufficient smoothness of the wavenumber functions.

\subsection{Stagnation criteria for the calculation of each atom/term}

After computing the wavenumber correction, the process restarts by evaluating a new amplitude and distance for the current $i$-th term of the decomposition. The iterations continue until a prescribed stagnation criterion is reached. This stagnation metric is defined as:
\begin{equation}
\epsilon_{i}^{[\ell]} =  \frac{ \epsilon_{k,i}^{[\ell]} + \epsilon_{d,i}^{[\ell]} + \epsilon_{\alpha,i}^{[\ell]} }{3}
\end{equation}
where:
\begin{equation}
\epsilon_{k,i}^{[\ell]} = \frac{\normi{k_{i}^{[\ell]}(\omega) - k_{i}^{[\ell-1]}(\omega)}_{\DT_{\omega}}}{\normi{k_{i}^{[\ell-1]}(\omega)}_{\DT_{\omega}}}
\end{equation}
\begin{equation}
\epsilon_{d,i}^{[\ell]} = \frac{ \left| d_{i}^{[\ell]} - d_{i}^{[\ell-1]} \right| }{ \left| d_{i}^{[\ell-1] } \right| }
\end{equation}
and
\begin{equation}
\epsilon_{\alpha,i}^{[\ell]} = \frac{ \left| \alpha_{i}^{[\ell]} - \alpha_{i}^{[\ell-1]} \right| }{\left| \alpha_{i}^{[\ell-1]} \right| }
\end{equation}
with $\left| \bullet \right|$ the absolute value of $\bullet$.

\subsection{Convergence criteria of the method}\label{sec:error}

New terms are added to the decomposition until a given approximation error or a maximum number of terms $N$ is reached. This error is defined as follows:
\begin{equation}
\xi = 100 \times \frac{\normi{ \hat{s}(\omega) -  \left[ \displaystyle \sum_{i=1}^{N} \alpha_i \hat{x}(\omega) e^{-j k_i(\omega) d_i} \right]}_{\DT_\omega}}{\normi{ \hat{s}(\omega)}_{\DT_\omega}}
\end{equation}

\subsection{Overview of the method}

The PISACMP algorithm proceeds iteratively, extracting one wave atom at a time from the residual signal. At each iteration, the propagation distance $d_i$, amplitude $\alpha_i$, and wavenumber dispersion curve $k_i(\omega)$ are jointly identified through a nested optimization loop. The inner loop refines the atom parameters until convergence, while the outer loop continues extracting atoms until the global reconstruction error falls below the prescribed tolerance $\varepsilon$. Algorithm~\ref{alg:PISACMP} summarizes the full procedure.

\begin{algorithm}[H]
\caption{Physics-Informed Single Atom Convolutional Matching Pursuit (PISACMP).}
\label{alg:PISACMP}
\begin{algorithmic}[1]

\Require Measured signal $\hat{s}(\omega)$, excitation $\hat{x}(\omega)$, tolerance $\varepsilon$, maximum number of atoms $N$
\Ensure Identified parameters $\{ \alpha_i, d_i, k_i(\omega) \}_{i=1}^{N}$

\State Initialize residual $\hat{s}_r^{[0]}(\omega) \gets \hat{s}(\omega)$

\For{$i = 1$ to $N$} \Comment{Outer loop: atom extraction}

    \State $k_i^{[0]}(\omega) \gets k_{S_0}(\omega)$ \Comment{Initialize with $S_0$ dispersion curve (Eq.~\eqref{eq:init_k})}
    \State $\ell \gets 1$

    \Repeat \Comment{Inner loop: joint parameter refinement}
        \State $d_i^{[\ell]} \gets \arg\min_{d} \; z_i(d)$ \Comment{Optimize propagation distance}
        \State $\alpha_i^{[\ell]} \gets \alpha_i\!\left(d_i^{[\ell]}\right)$ \Comment{Update amplitude}
        \State Solve $\SDtens{M}_i^{[\ell]} \, \Delta \SDvect{\beta}_i^{[\ell]} = \SDvect{f}_i^{[\ell]}$ \Comment{Update wavenumber corrections}
        \State $k_i^{[\ell]}(\omega) \gets k_i^{[\ell-1]}(\omega) + \Delta k_i^{[\ell]}(\omega)$
        \State $\epsilon_i^{[\ell]} \gets \dfrac{\epsilon_{k,i}^{[\ell]} + \epsilon_{d,i}^{[\ell]} + \epsilon_{\alpha,i}^{[\ell]}}{3}$ \Comment{Averaged convergence criterion}
        \State $\ell \gets \ell + 1$
    \Until{$\epsilon_i^{[\ell]} < \varepsilon$}

    \State Store converged atom: $\alpha_i \gets \alpha_i^{[\ell]}$,\; $d_i \gets d_i^{[\ell]}$,\; $k_i(\omega) \gets k_i^{[\ell]}(\omega)$
    \State Update residual: $\hat{s}_r^{[i]}(\omega) \gets \hat{s}_r^{[i-1]}(\omega) - \alpha_i\, \hat{x}(\omega)\, e^{-j k_i(\omega) d_i}$
    \State Compute global reconstruction error $\xi$
    \If{$\xi < \varepsilon$} \Comment{Early stopping criterion}
        \State \textbf{break}
    \EndIf

\EndFor

\State \Return $\{ \alpha_i,\, d_i,\, k_i(\omega) \}_{i=1}^{N}$

\end{algorithmic}
\end{algorithm}

\section{Damage Localization using PISACMP via Path Length Estimation}\label{sec:PISACMP_dam_loc}

The decomposition introduced in the previous Section provides estimates of the effective propagation distances associated with the scattered wavefield. 
These quantities represent the total travel path of different waves that propagates in the structure. Among them, it exists the one related to the wave travelled between from the actuator to damage location and from it to each sensor. 

However, here the main difficulty consists in determining the distances travelled by the waves such as they give us important information about the location of damage. 

Furthermore, the distances obtained from the phase-based formulation are not directly expressed in physical units. This limitation arises from the intrinsic scale indeterminacy of the exponential term in the proposed propagation model \eqref{eq:phy_wave_prop}. 
Indeed, since only the product $k_i(\omega)d_i$ appears inside the complex exponential, infinitely many equivalent pairs $(k_i(\omega), d_i)$ generate the same response. 
More precisely, for any scalar $c \in \mathbb{R} \setminus \{0\}$:
\begin{equation}
k_i(\omega) \rightarrow \frac{k_i(\omega)}{c},
\qquad
d_i \rightarrow c\, d_i,
\end{equation}
leaves the exponential term unchanged. 

Consequently, the distances identified by the decomposition are determined only up to a multiplicative constant.

In the following, a calibration strategy is introduced to resolve this ambiguity, allowing the estimated path lengths to be interpreted in physical units and directly related to the known geometry of the structure. 
Once properly scaled, these distances are used to determine the spatial position of the damage, as will be shown in the following sections.

\subsection{Forward Model and Physical Assumptions}

Consider a two-dimensional structure instrumented with a single actuator located at
$\mathbf{a} = (x_a, y_a) \in \mathbb{R}^2$
and $n$ sensors located at
$\mathbf{s}_i = (x_i, y_i)$, $i = 1,\dots,n$.
An unknown damage (scattering point) is located at
$\mathbf{x}_d = (x_d, y_d)$.

Under the single-scattering assumption, the total propagation distance associated with sensor $i$ satisfies:
\begin{equation}
\label{eq:dam_det_eq_revised}
D_i = \|\mathbf{x}_d - \mathbf{a}\|_2 + \|\mathbf{x}_d - \mathbf{s}_i\|_2
\end{equation}

The first term represents propagation from the actuator to the damage, and the second term represents propagation from the damage to sensor $i$. 

Therefore, the knowledge of $D_i$ for $i\in (1, ..., n)$ allows the damage location to be determined from geometric considerations.

\subsection{Scale Calibration via Direct Propagation Path}

To correctly obtains distances $D_i$ to be used in \eqref{eq:dam_det_eq_revised}, one must convert the numerically estimated distances into physical units. 

To solve this problem, here, the direct actuator-to-sensor propagation path is used as a reference. 
For each sensor $i$, the true geometric distance is:
\begin{equation}
D_{a,s_i}^{\text{real}} = \|\mathbf{a} - \mathbf{s}_i\|_2
\end{equation}

By applying the PISACMP decomposition to the undamaged structural response and determining an estimated distance $d_{\text{ref}}$ corresponding to the direct wave arrival, a global scaling factor can then be defined:
\begin{equation}
\beta = \frac{D_{a,s_i}^{\text{real}}}{d_{\text{ref}}}
\end{equation}

Such as all estimated distances can be corrected according to:
\begin{equation}
D_i^{\text{corr}} = \beta \, d_i
\end{equation}

This transformation ensures that the reconstructed path lengths are expressed in physical units and are consistent with the known geometry.

However, the challenge that remains here, corresponds to the correct numerical determination of the first wave-packet arrival at sensor. This is a real challenge, since the first arrival wave-packet can be very small with respect to the rest of the measured signal at sensor location.

In this context, in the following section, a new methodology is proposed in order to adapt PISACMP method shown in Section \ref{sec:PIMPM} to determine this first arrival packet.

\subsection{Distance determination related to first arrivals wave-packets}\label{sec:dist_first_arrival}

The main idea here corresponds to determine numerically the distance traveled from excitation to sensors as well as the distance from excitation to damage and damage to sensors. When performing SHM, those distances corresponds to the first wave packets that arrives at sensors location, as well as for the non damaged structure and the damaged ones.

\begin{figure}[H]
\centering
\includegraphics[width=0.7\textwidth]{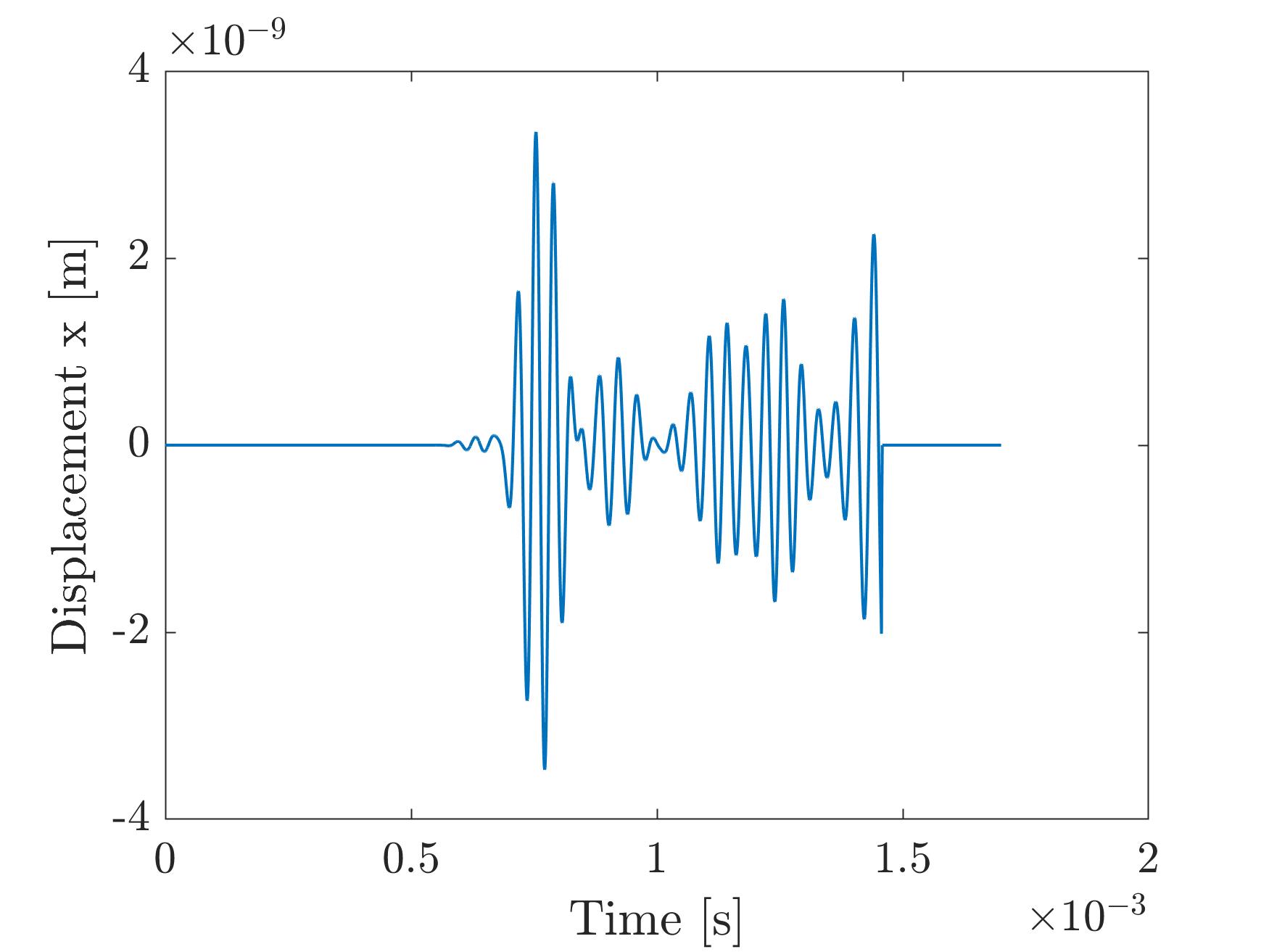}
\caption{Complex situations in where determining the first wave-packet is not evident, here the first arrival atom is mark as a red rectangle.}
\label{fig:dam_challenge_signal}
\end{figure}

Indeed, the determination of first arrival packets is a challenge problem, since in many situation, those packets have negligible amplitude with respect to the overall measured signal. To illustrate this, lets consider Fig.~\ref{fig:dam_challenge_signal}, here one can clearly see that the first arrival wave-packet is considerably small with respect to the overall signal. In this regard, if we follows the optimal distance determination presented in Section \ref{sec:amp_distance_PIMPM}, the algorithm will not see this atom easily, since its contribution to the overall approximation of the signal is small. 

To overcome this limitation and adapt PISACMP to SHM applications, this subsection proposes addressing the determination of these distances by slightly modifying the error criterion minimized by PISACMP, thereby enabling the direct determination of the first atom corresponding to the first arrival packet.

\begin{figure}[H] 
\centering
\begin{subfigure}{0.49\textwidth}
\includegraphics[width=\textwidth]{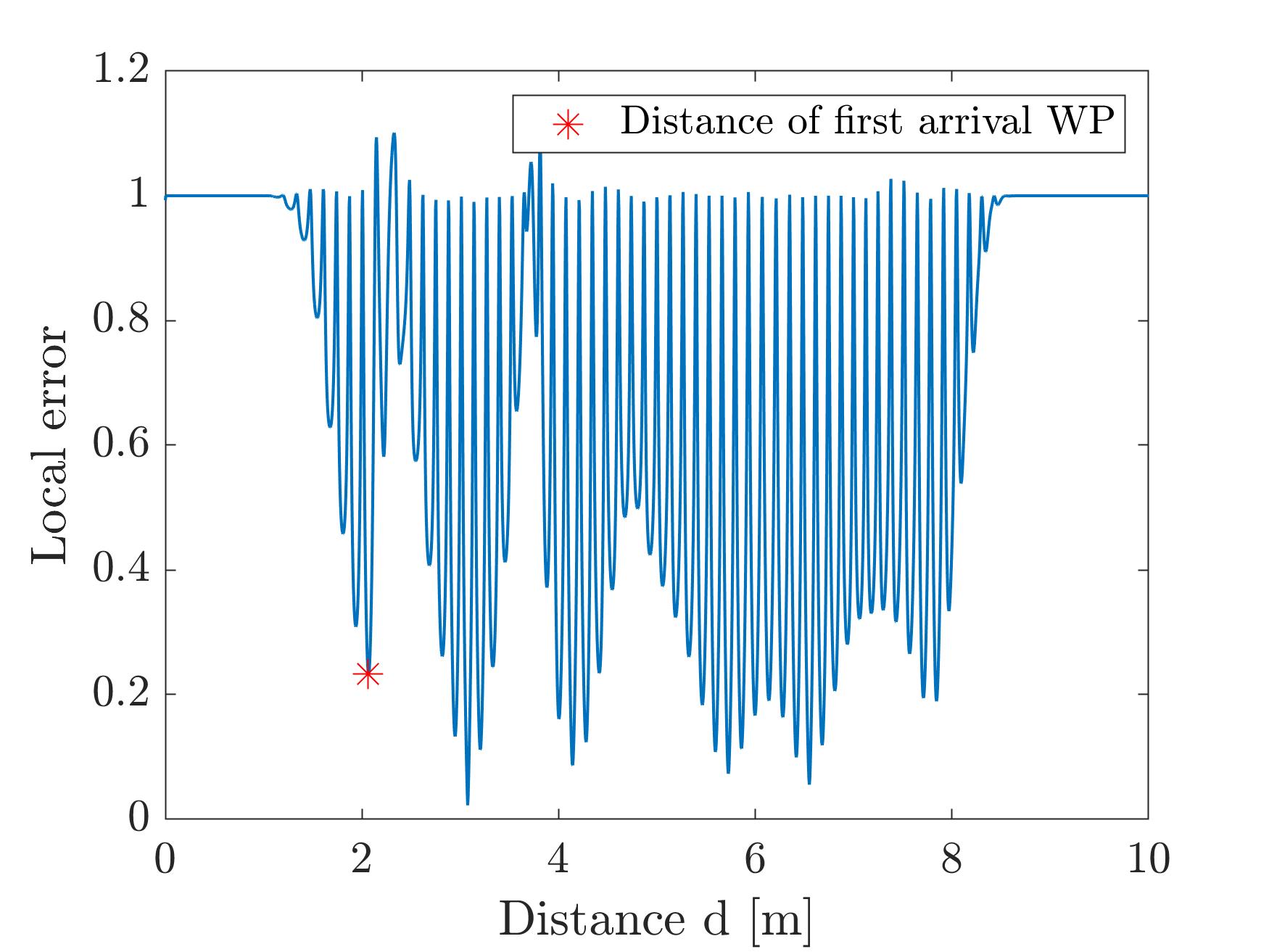}
\caption{Original local error.}
\label{fig:Local_error}
\end{subfigure}
\begin{subfigure}{0.49\textwidth}
\includegraphics[width=\textwidth]{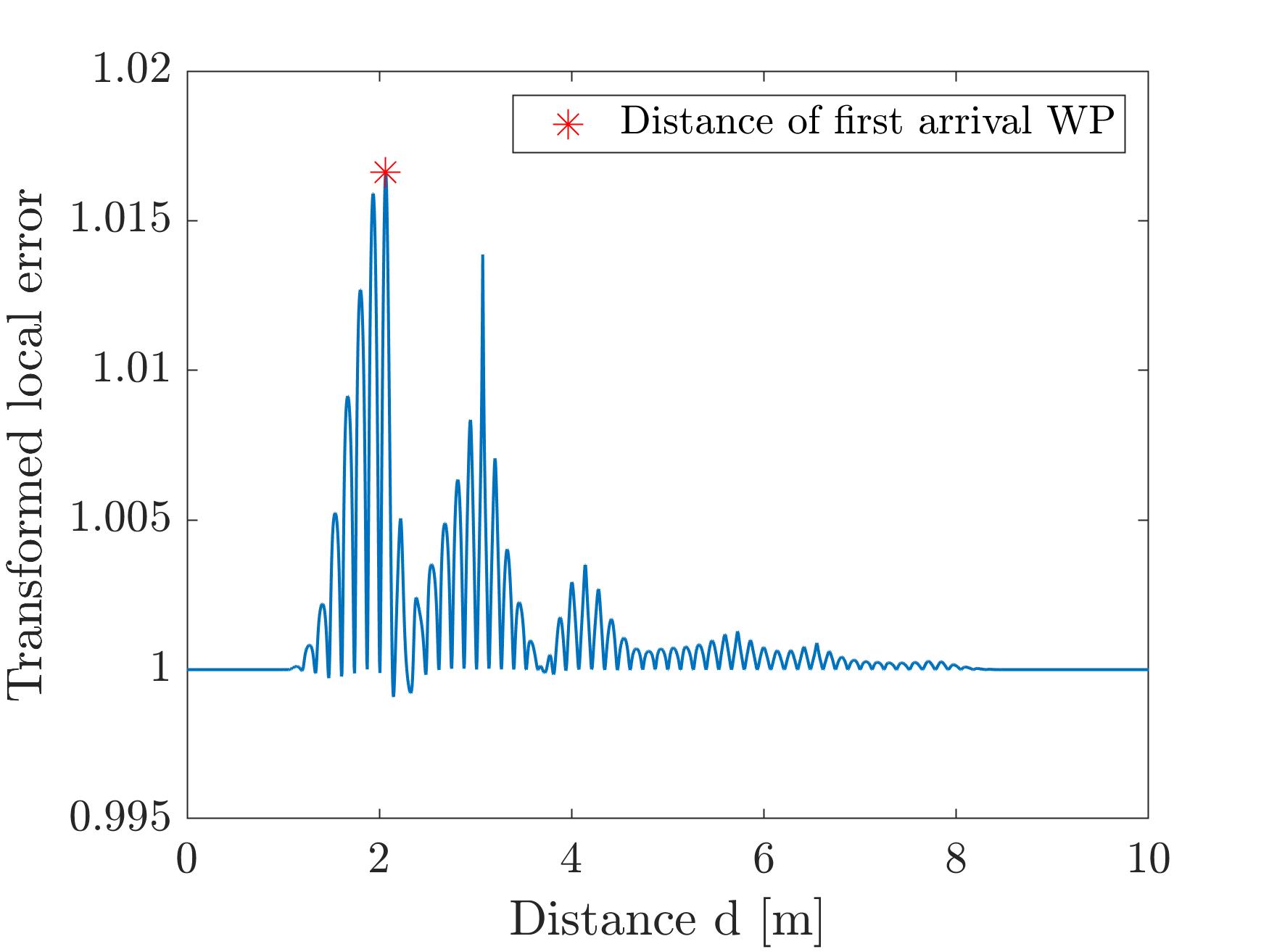}
\caption{Transformed local error.}
\label{fig:Transformed_local_error}
\end{subfigure}
\caption{Local and transformed local errors used to determine first wake-packet arrival.}
\end{figure}

\begin{figure}[H]
\centering
\includegraphics[width=0.7\textwidth]{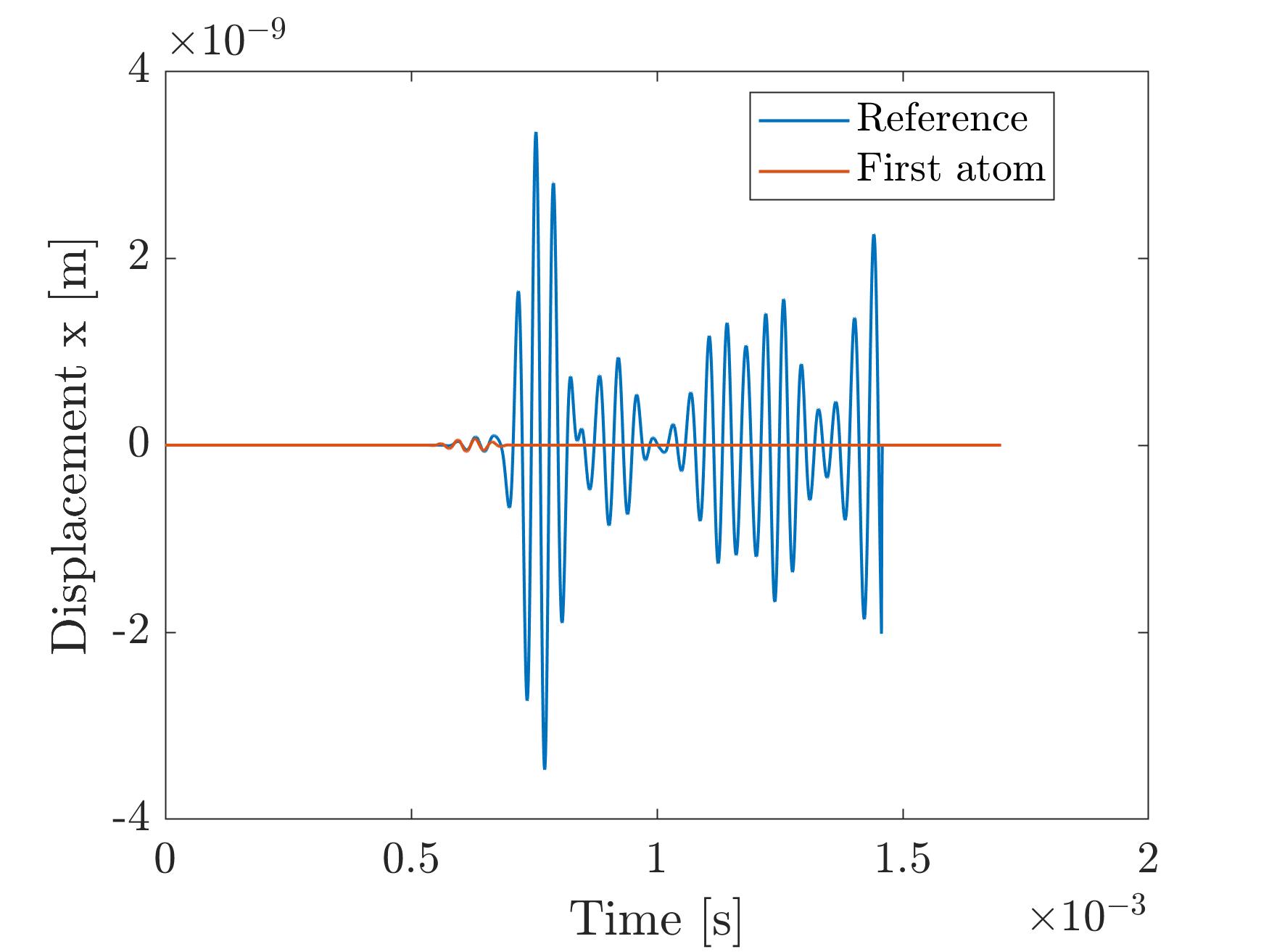}
\caption{Challenge determination of first arrival atom, the proposed strategy works without problem.}
\label{fig:dam_challenge_atom}
\end{figure}

By defining an optimized atom at distance $d$ as:
\begin{equation}
s_{\text{atom}}(d;t) = \mathcal{F}^{-1}( \alpha_i(d) \hat{x}(\omega) e^{-j k_i(\omega) d} )
\end{equation}
one can define a local error definition as follows:
\begin{equation}
\xi_{\text{local}}(d) = \frac{\norm{ s(t) s_{\text{atom}}(d;t) - s_{\text{atom}}(d;t)^{2} }_{\DT}}{  \norm{ s(t) s_{\text{atom}}(d;t) +  s_{\text{atom}}(d;t)^{2} }_{\DT} }
\end{equation}
%
%

%
%
with:
\begin{equation}
\norm{\bullet}^{2}_{\DT} = \int_{0}^{T} (\bullet)^T(\bullet)dt
\end{equation}
This local error makes atoms comparable, regardless of their contribution on the overall signal approximation. However, the local error in function of distance $d$ can show many small errors as illustrated on Fig.~\ref{fig:Local_error}, each of one corresponding to an atom that correctly fit the reference signal on a localized temporal interval. In order to ensure that the determined distance corresponds to the first arrival packet, here one proposes the following transformation of the local error after it is computed:
\begin{equation}
\tilde{\xi}_{\text{local}}(d) = J(\xi_{\text{local}}(d))
\end{equation}
with:
\begin{equation}
J(\xi_{\text{local}}(d)) = \xi_{\text{local}}(d)^{-\frac{1}{(d+1)^4}}
\end{equation}
The considered function $J(\xi_{\text{local}}(d))$ allows to gives more importance to wave-packets that minimize the local error ensuring at the same time the smallest $d$ as possible. In this sense, the selected distance ensuring the first arrival is simply selected as:
\begin{equation}
d = \underset{ d }{\text{arg max}}( J(\xi_{\text{local}}(d)) )    
\end{equation}
This transformed local error of Fig.~\ref{fig:Local_error} is presented in Fig.~\ref{fig:Transformed_local_error}.
Figure \ref{fig:dam_challenge_atom} shows the atom determined following the presented strategy. One can see that the first arrival is correctly determined.

\subsection{Damage Localization Framework}\label{sec:dam_loc_framework}

After calibration, the corrected distances satisfy:
\begin{equation}
\label{eq:ellipse_revised}
D_i^{\text{corr}} = 
\|\mathbf{x}_d - \mathbf{a}\|_2 
+ 
\|\mathbf{x}_d - \mathbf{s}_i\|_2
\end{equation}
The corrected distance $D_i^{\text{corr}}$ is obtained by applying the PISACMP decomposition and determining the first wave-packet following Section \ref{sec:dist_first_arrival} of the scattered component of the signal, computed as the difference between the damaged and undamaged responses:
\begin{equation}
s_i^{\text{dam}}(t) - s_i^{\text{undam}}(t)
\approx
\mathcal{F}^{-1}
\left[
\alpha_1 \hat{x}(\omega)
e^{-j k_1(\omega) D_i^{\text{corr}}}
\right]
\end{equation}

For $n \ge 3$ non-collinear sensors, the damage location is uniquely determined in the noise-free case. 
In practice, measurement noise prevents exact satisfaction of \eqref{eq:ellipse_revised}, therefore, one must solve in a least square way, which is robust to noise. Here, since the problem is nonlinear, the nonlinear least squares problem is considered for the damage location determination.

In this sense, the norm to be solve  is considered as follows: 
\begin{equation}
\label{eq:nonlinear_problem_revised}
\min_{\mathbf{x}_d \in \mathbb{R}^2}
\sum_{i=1}^{n}
\left(
\|\mathbf{x}_d - \mathbf{a}\|_2
+
\|\mathbf{x}_d - \mathbf{s}_i\|_2
-
D_i^{\text{corr}}
\right)^2
\end{equation}
Problem \eqref{eq:nonlinear_problem_revised} is smooth and can be efficiently solved using Gauss–Newton or Levenberg–Marquardt algorithms.

This procedure establishes a direct and physically consistent connection between the dispersion-aware signal decomposition and the geometric localization of structural damage, following a standard \emph{Elliptical Localization} approach.
\begin{figure}[H]
\centering
\includegraphics[width=0.4\textwidth]{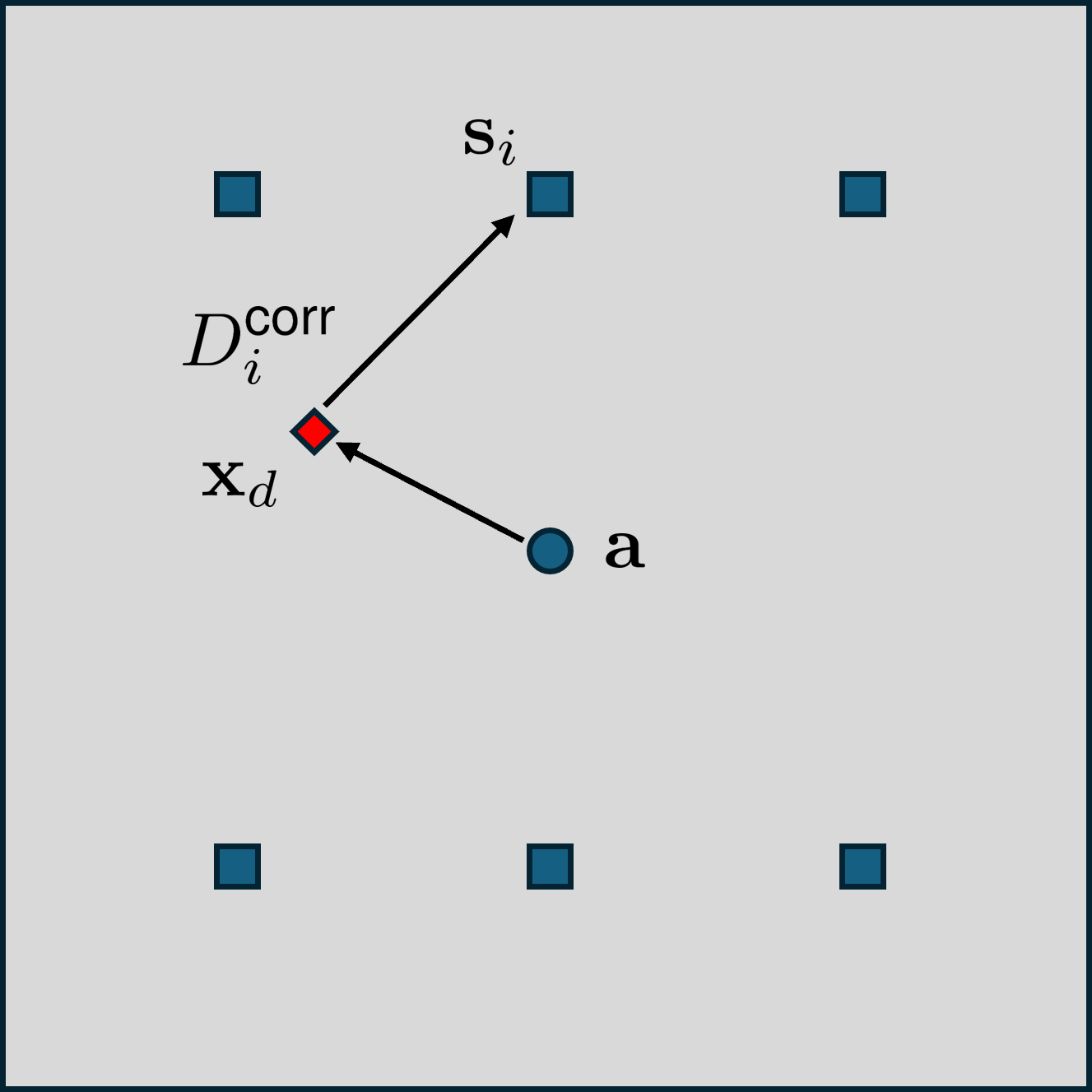}
\caption{Reference problem.}\label{fig:Ref_prob}
\end{figure}
\textbf{Remark}: It should be noted that the proposed damage localization strategy presented here is applicable only to planar surfaces, since it relies on the Euclidean distance. For curved surfaces, the notion of distance traveled on a plane must be extended by considering geodesic distances along the surface. This interesting extension is left as a perspective for future work.

\section{Numerical results}\label{sec:numerical_results}

To illustrate the performance of the proposed method, here one separate the results into two Sections. At one hand, Section \ref{sec:1D_PISACMP_approx} corresponds to the approximation of temporal signals, and at the other hand, Section \ref{sec:PISACMP_dam_det} considers the PISACMP method to perform damage detection on a 2D thin-plate.

\subsection{PISACMP for the approximation of 1D signals}\label{sec:1D_PISACMP_approx}

\subsubsection{Simple signal approximation: infinite isotropic plate}\label{sec:simple_signal_approx}

\begin{figure}[H] 
\centering
\begin{subfigure}{0.49\textwidth}
\includegraphics[width=\textwidth]{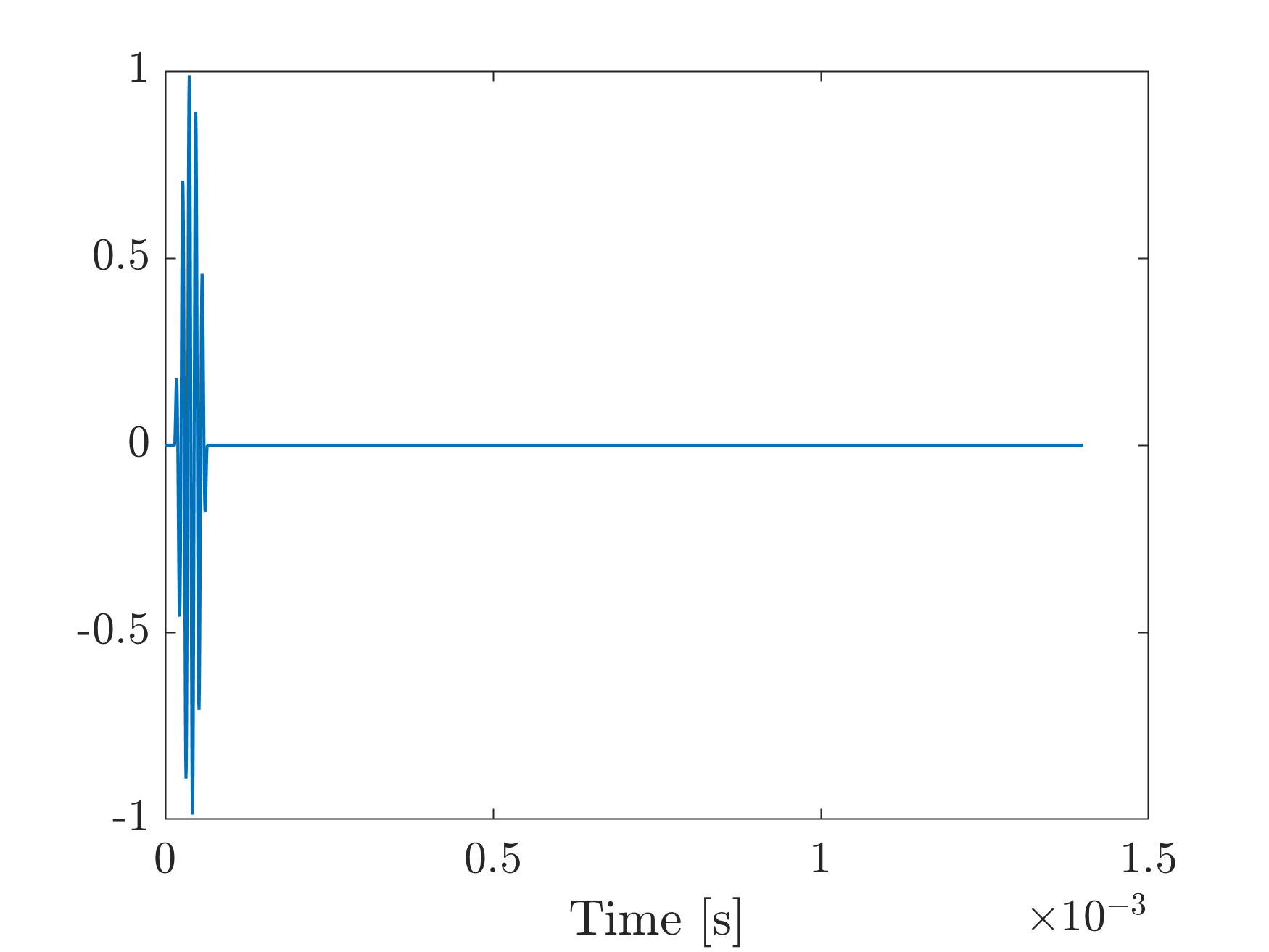}
\caption{Input signal.}
\label{fig:Input_simple}
\end{subfigure}
\begin{subfigure}{0.49\textwidth}
\includegraphics[width=\textwidth]{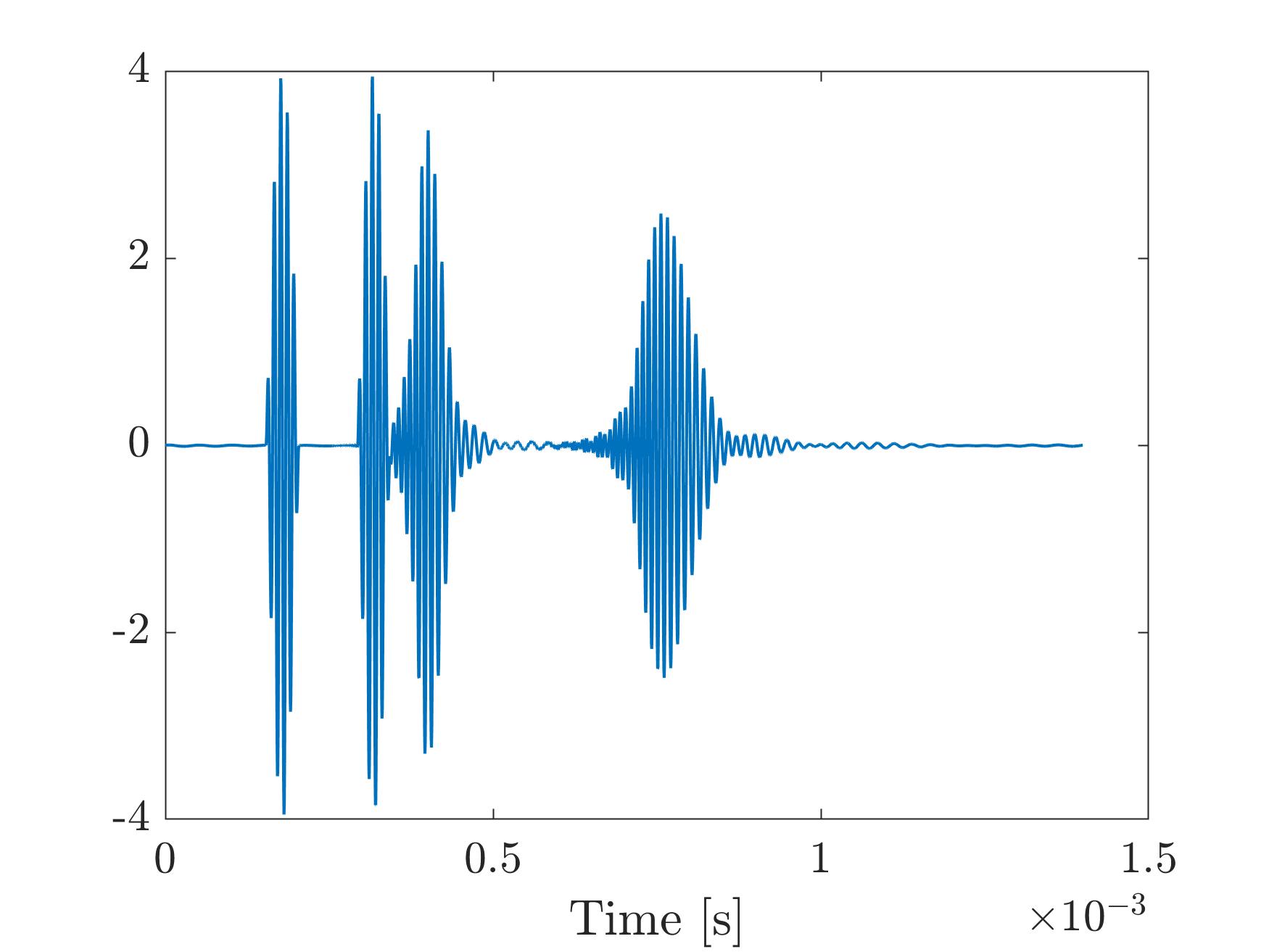}
\caption{Reference signal.}
\label{fig:Reference_sig_simple}
\end{subfigure}
\caption{Input signal (Fig.~\ref{fig:Input_simple}) and reference signal to be approximated (Fig.~\ref{fig:Reference_sig_simple}).}
\end{figure}

\begin{figure}[H]
\centering
\includegraphics[width=0.6\textwidth]{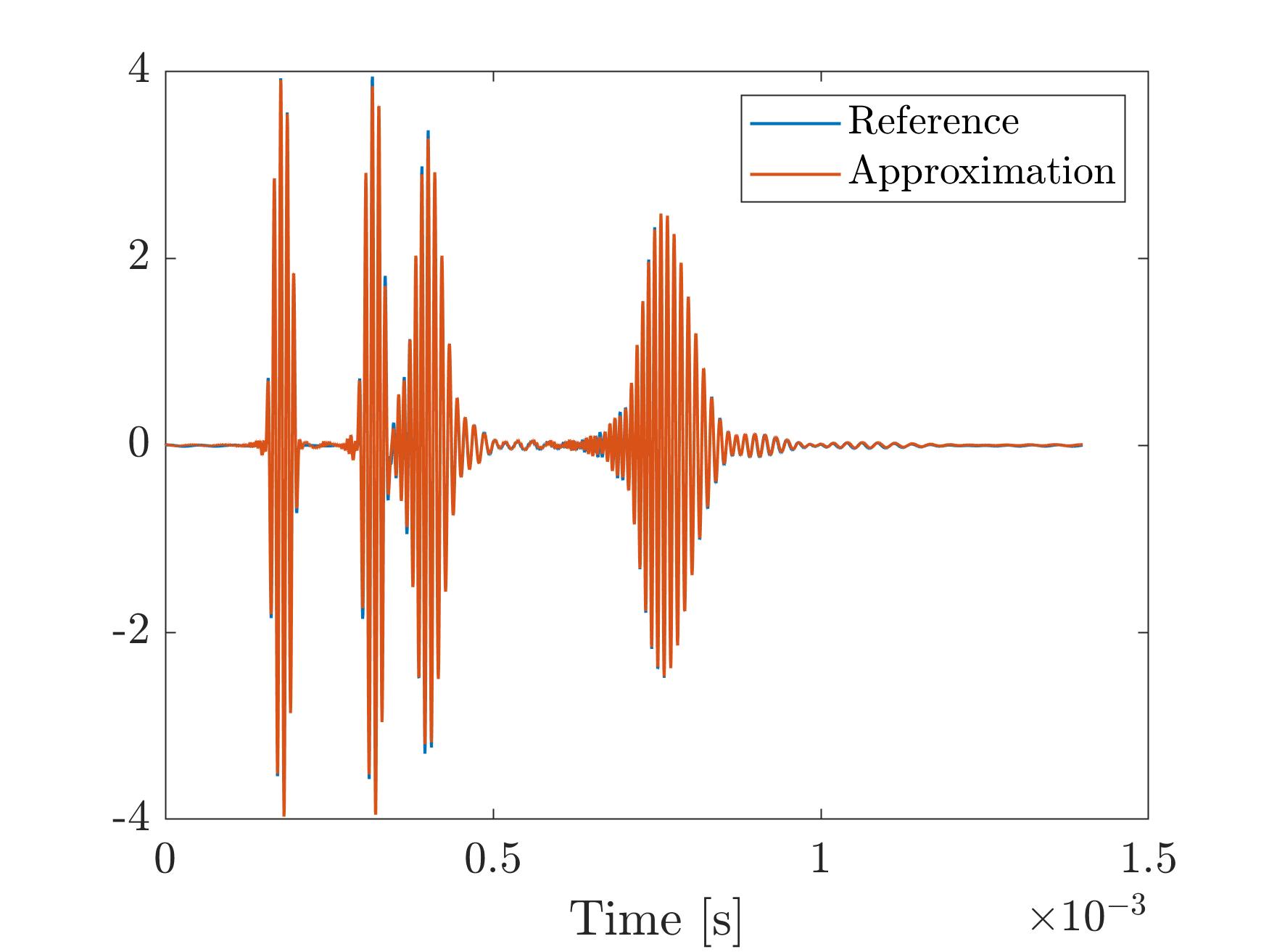}
\caption{Comparison of reference and approximation.}\label{fig:comparison_simple}
\end{figure}

\begin{figure}[H]
\centering
\includegraphics[width=0.6\textwidth]{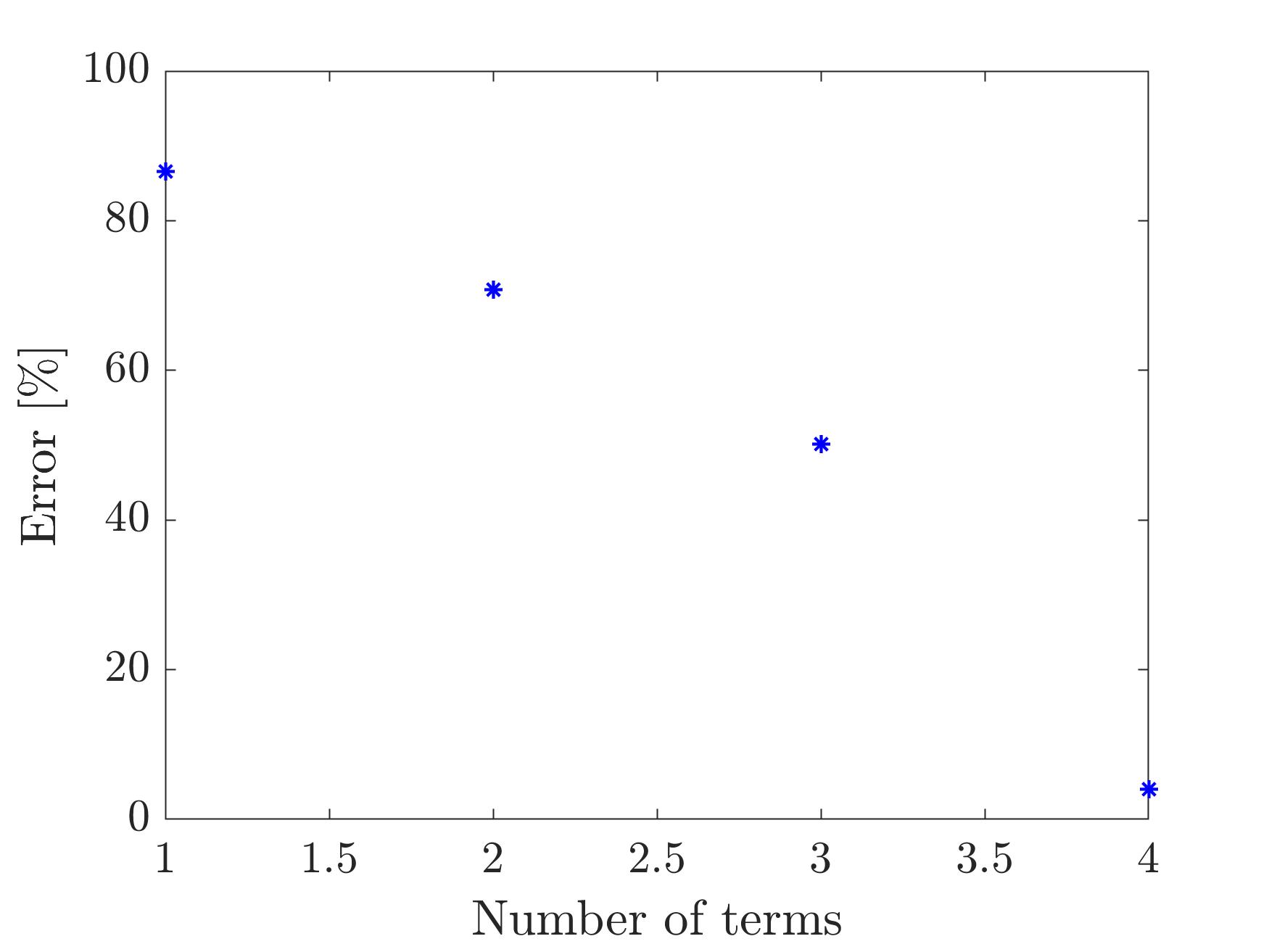}
\caption{Error versus number of submodes of the decomposition.}\label{fig:Error_simple}
\end{figure}

Here, an academic example is considered, consisting of a signal generated from analytical expressions describing wave propagation in an isotropic plate. The reference signal to be approximated is shown in Fig.~\ref{fig:Reference_sig_simple}. This signal is constructed using Eqs.~\eqref{eq:wav_non_disp}, \eqref{eq:wav_disp}, and \eqref{eq:phy_wave_prop_analytical}, considering a plate with thickness $h = 2$~[mm], Young's modulus $E = 70$~[GPa], Poisson's ratio $\nu = 0.3$, and density $\rho = 1500$~[kg/m$^3$]. The excitation signal corresponds to the input shown in Fig.~\ref{fig:Input_simple}. Two propagation distances, $d = 1$~[m] and $d = 2$~[m], are considered for both non-dispersive and dispersive wave modes. The input signal consists of a burst excitation with $5$ cycles, centered at $\omega = 0.62 \times 10^6$~[rad/s].
%
%
%
%
After applying PISACMP approximation \eqref{eq:phy_wave_prop} using $6$ Chebyshev shape functions, we obtain a reconstruction error of $4 \%$ (by using the error definition of Section \ref{sec:error}), the approximation compared with the reference signal is shown in Fig.~\ref{fig:comparison_simple}.
The curve of error versus the number of submodes of the decomposition is illustrated in Fig.~\ref{fig:Error_simple}.
%
%

%

\begin{figure}[H] 
\centering
\begin{subfigure}{0.49\textwidth}
\includegraphics[width=\textwidth]{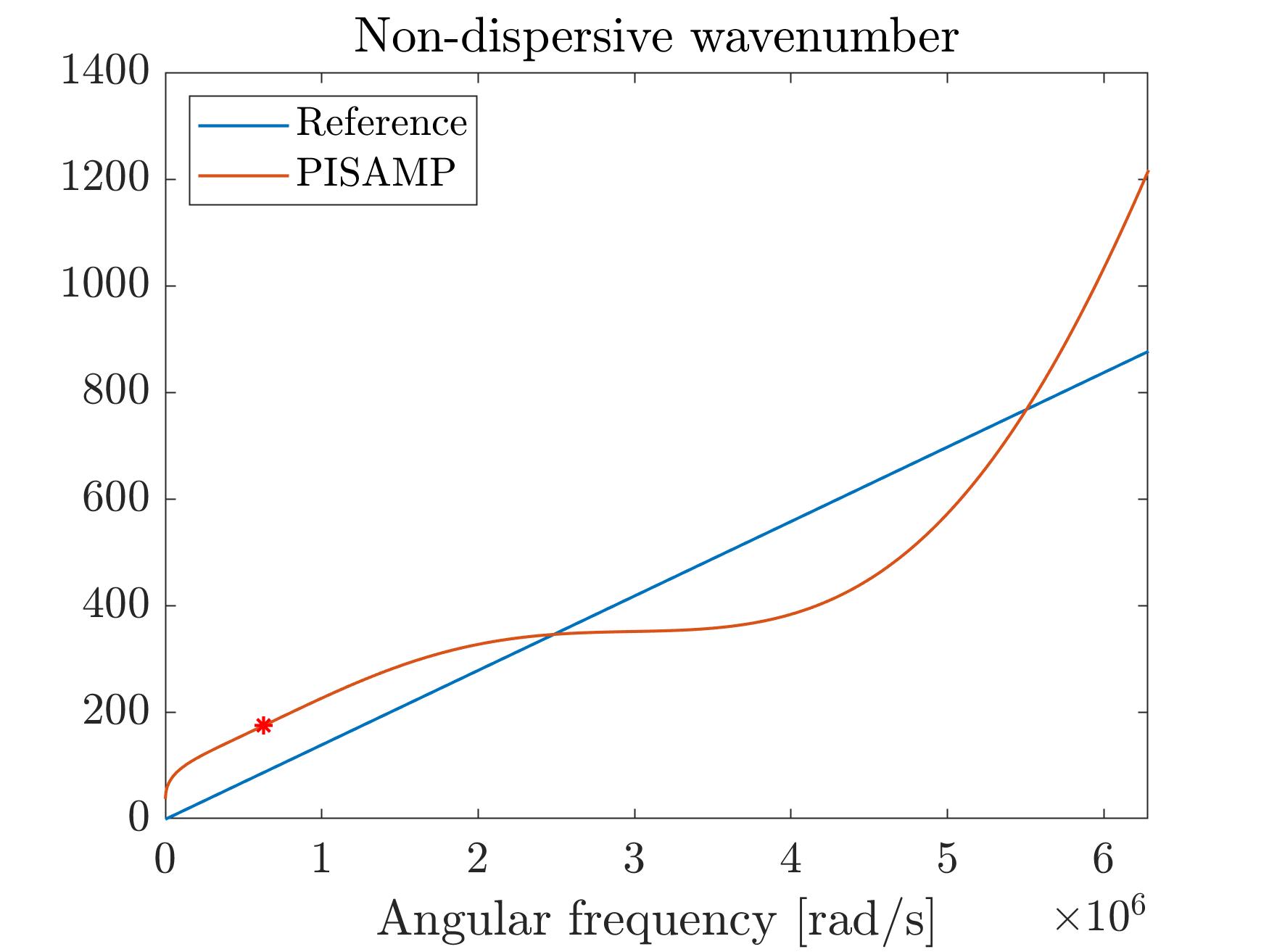}
\caption{Comparison of non-dispersive wavenumber.}
\label{fig:Non_Disp_wavenumber}
\end{subfigure}
\begin{subfigure}{0.49\textwidth}
\includegraphics[width=\textwidth]{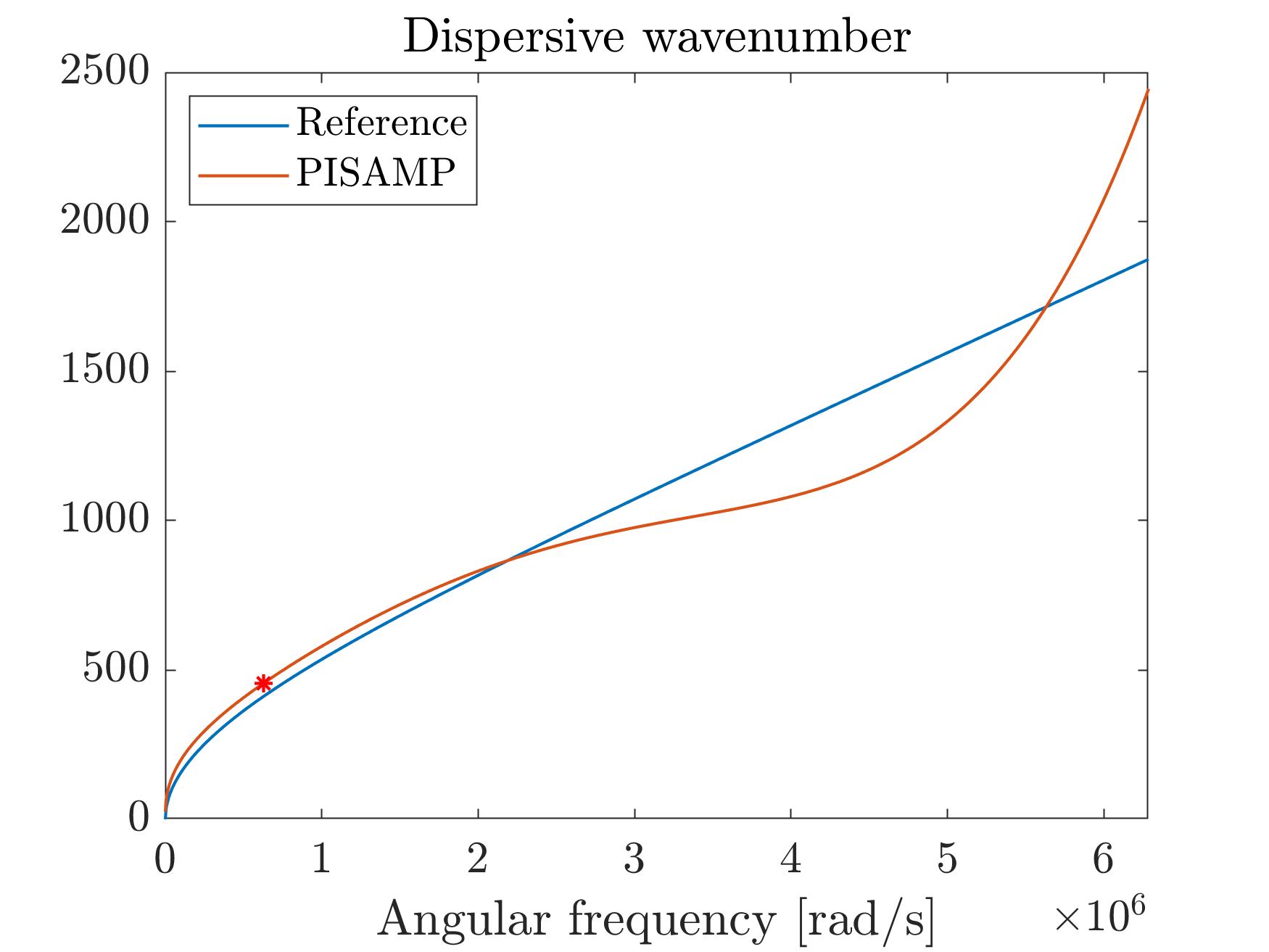}
\caption{Comparison of Dispersive wavenumber.}
\label{fig:Disp_wavenumber}
\end{subfigure}
\caption{Reference and PISACMP Wavenumber functions obtained, the red dot indicates the input central burst frequency.}
\label{fig:Comp_Wavenumbers_PISACMP}
\end{figure}

As stated in previous Sections, the PISACMP compute wavenumber functions as well as distances but that can have a different amplitude with respect to the reference one. In this context, if one scale the amplitude of the wavenumbers by matching the distances with the reference ones, one can check that the proposed decomposition actually obtained a good physical approximation. The wavenumber functions obtained by PISACMP for the dispersive and non-dispersive modes as well as the one used to construct the approximated signal are illustrated in Fig.~\ref{fig:Comp_Wavenumbers_PISACMP}.

\begin{figure}[H]
\centering
\includegraphics[width=0.7\textwidth]{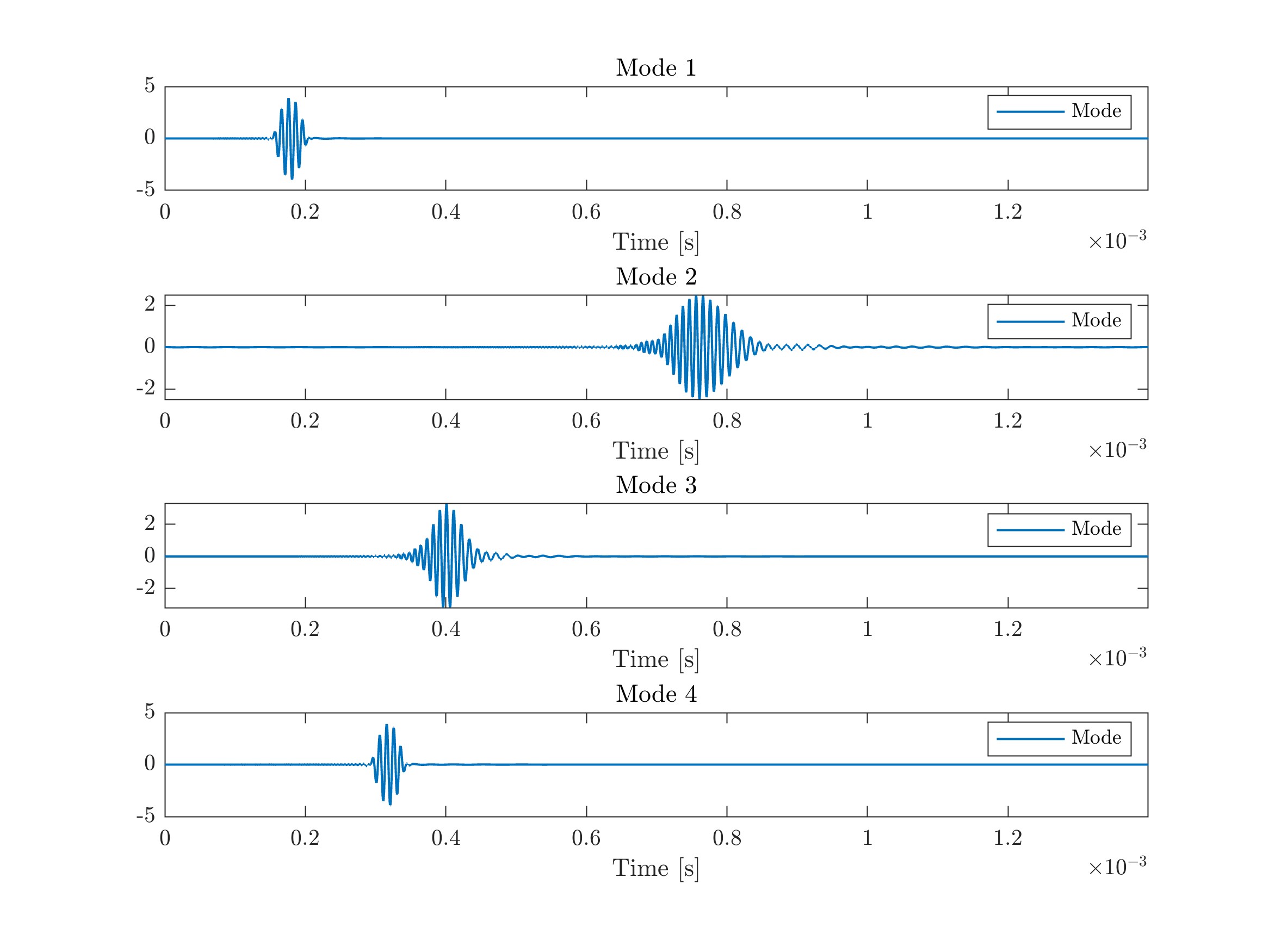}
\caption{Submodes of the decomposition.}\label{fig:submodes_simple}
\end{figure}

Finally, Fig.~\ref{fig:submodes_simple} shows the submodes obtained through the proposed decomposition.

As can be seen from Fig.~\ref{fig:submodes_simple}, the proposed signal approximation method allows to successfully identify the modes corresponding to a non-dispersive propagation (such as submodes $1$ and $4$) from the dispersive ones (submodes $2$ and $3$), due to the high physical content in the employed approximation.

\subsubsection{Complex signal approximation: fan cowl of the A380 Fan Cowl Structure}\label{sec:complex_signal_approx}

\begin{figure}[!ht]
\centering
\includegraphics[height=4.5cm]{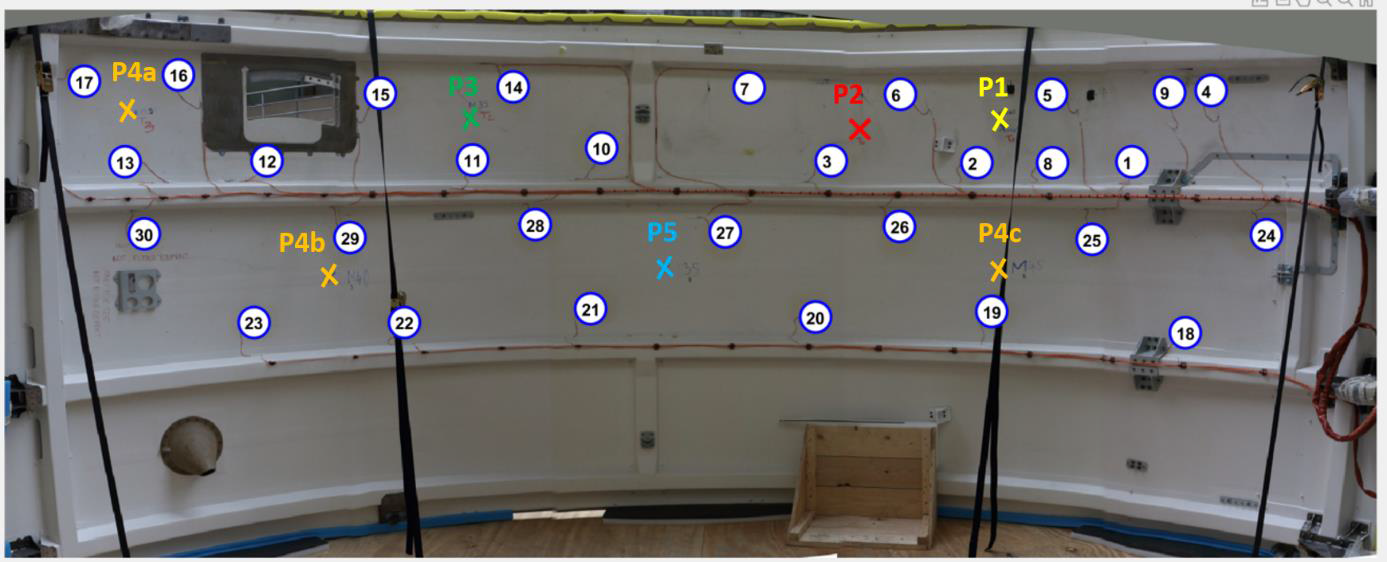}
    \caption{Overview of the geometrical configuration of the Fan Cowl Structure (FC) experimental datasets.}
    \label{fig:FC}
\end{figure}

\begin{figure}[!ht]
\centering
    \includegraphics[width=0.75\textwidth]{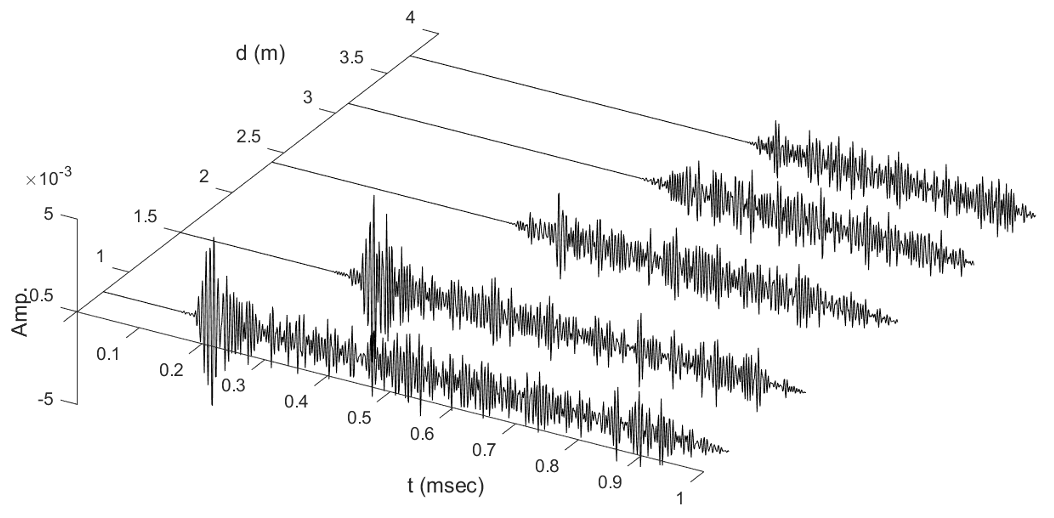}
    \caption{Experimental signals for various propagating distances.}
    \label{fig:XPsigs}
\end{figure}

Here, the proposed PISACMP is employed for signal approximation using data from an experimental set-up involving the fan cowl section of an Airbus A380 nacelle (see Fig.~\ref{fig:FC}).
The experimental signals considered in this work are presented in Fig.~\ref{fig:XPsigs}. As can be observed, these signals are more complex than the simulated ones discussed previously. In addition to non-dispersive and dispersive modes, reflections from structural boundaries and inhomogeneities—such as stiffeners (see Fig.~\ref{fig:FC})—generate additional wave packets that are also detected by the receiving piezoelectric sensors. Furthermore, due to material damping in composite structures, the amplitude of the first peak decreases as the propagation distance increases. This structure has been previously utilized by the authors in related studies; further details can be found in \citep{rebillat2018peaks, rebillat2020damage, Guo2022b,rodriguez2025single}.

%
%
\begin{figure}[H] 
\centering
\begin{subfigure}{0.49\textwidth}
\includegraphics[width=\textwidth]{figures/Input_signal_WM_simple.png}
\caption{Input signal.}
\label{fig:Input_complex}
\end{subfigure}
\begin{subfigure}{0.49\textwidth}
\includegraphics[width=\textwidth]{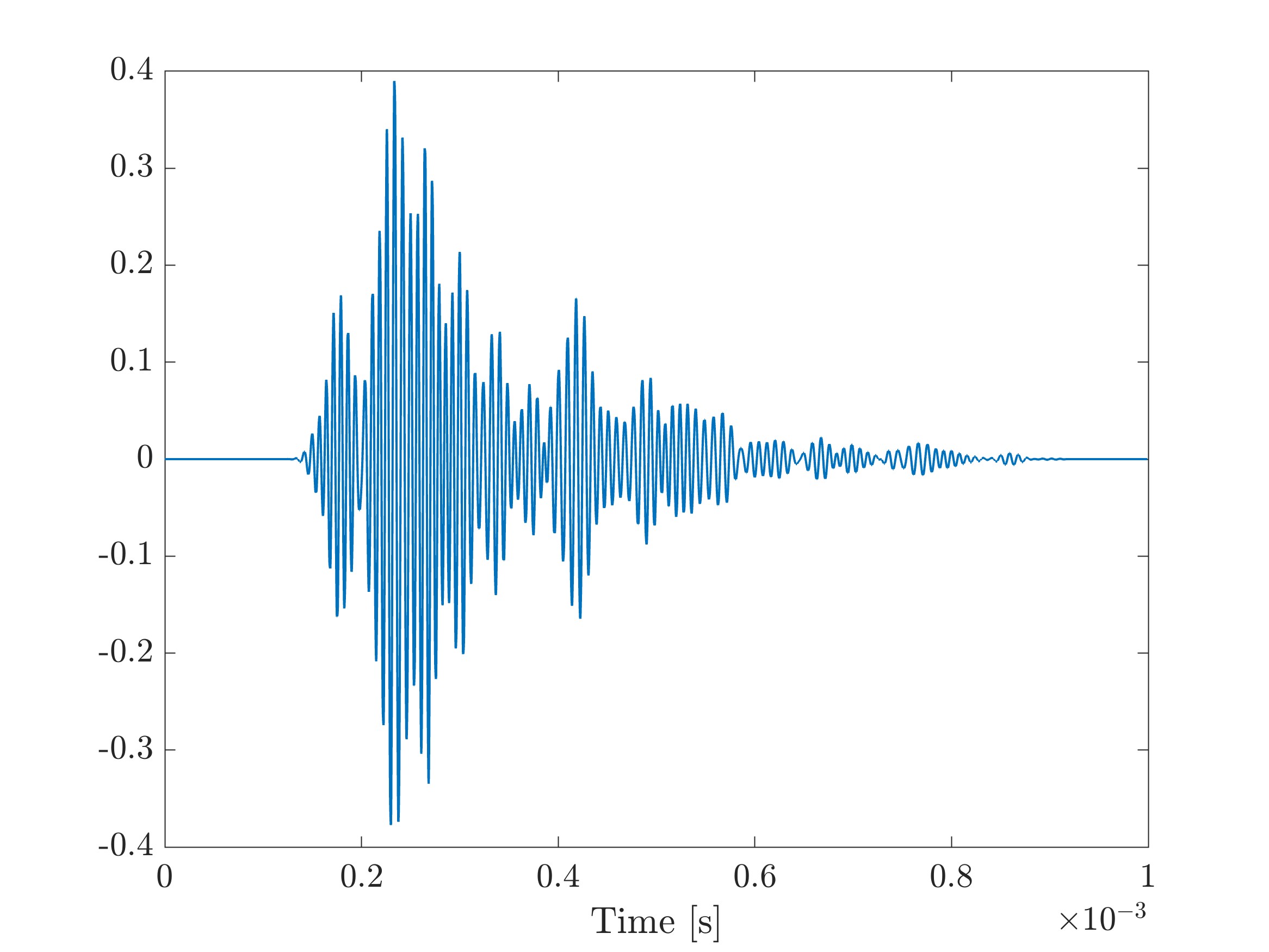}
\caption{Reference signal.}
\label{fig:Reference_sig_complex}
\end{subfigure}
\caption{Input signal (Fig.~\ref{fig:Input_complex}) and reference signal to be approximated (Fig.~\ref{fig:Reference_sig_complex}).}
\end{figure}

\begin{figure}[H]
\centering
\includegraphics[width=0.6\textwidth]{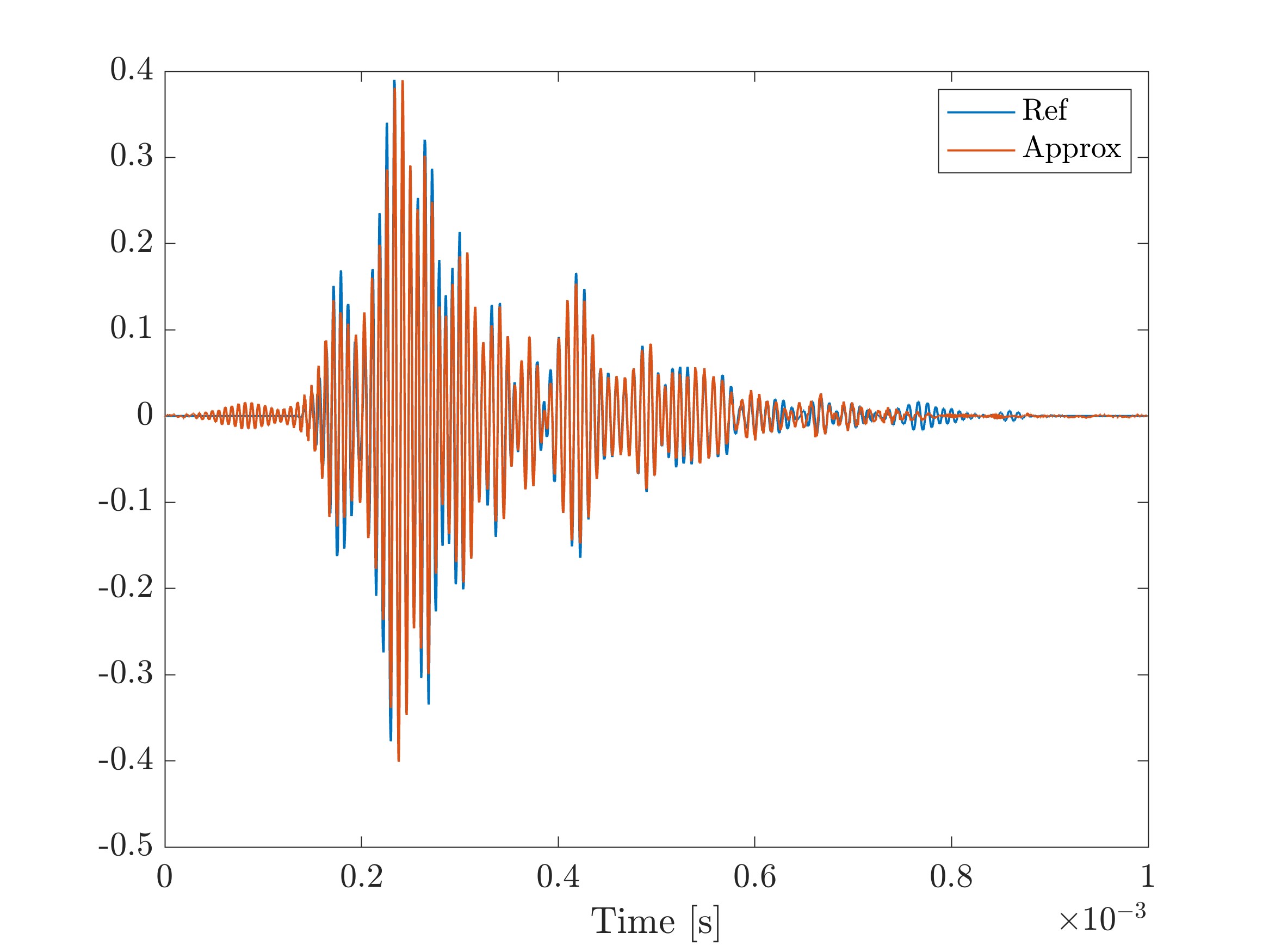}
\caption{Comparison of reference and approximation.}\label{fig:comparison_complex}
\end{figure}

\begin{figure}[H]
\centering
\includegraphics[width=0.6\textwidth]{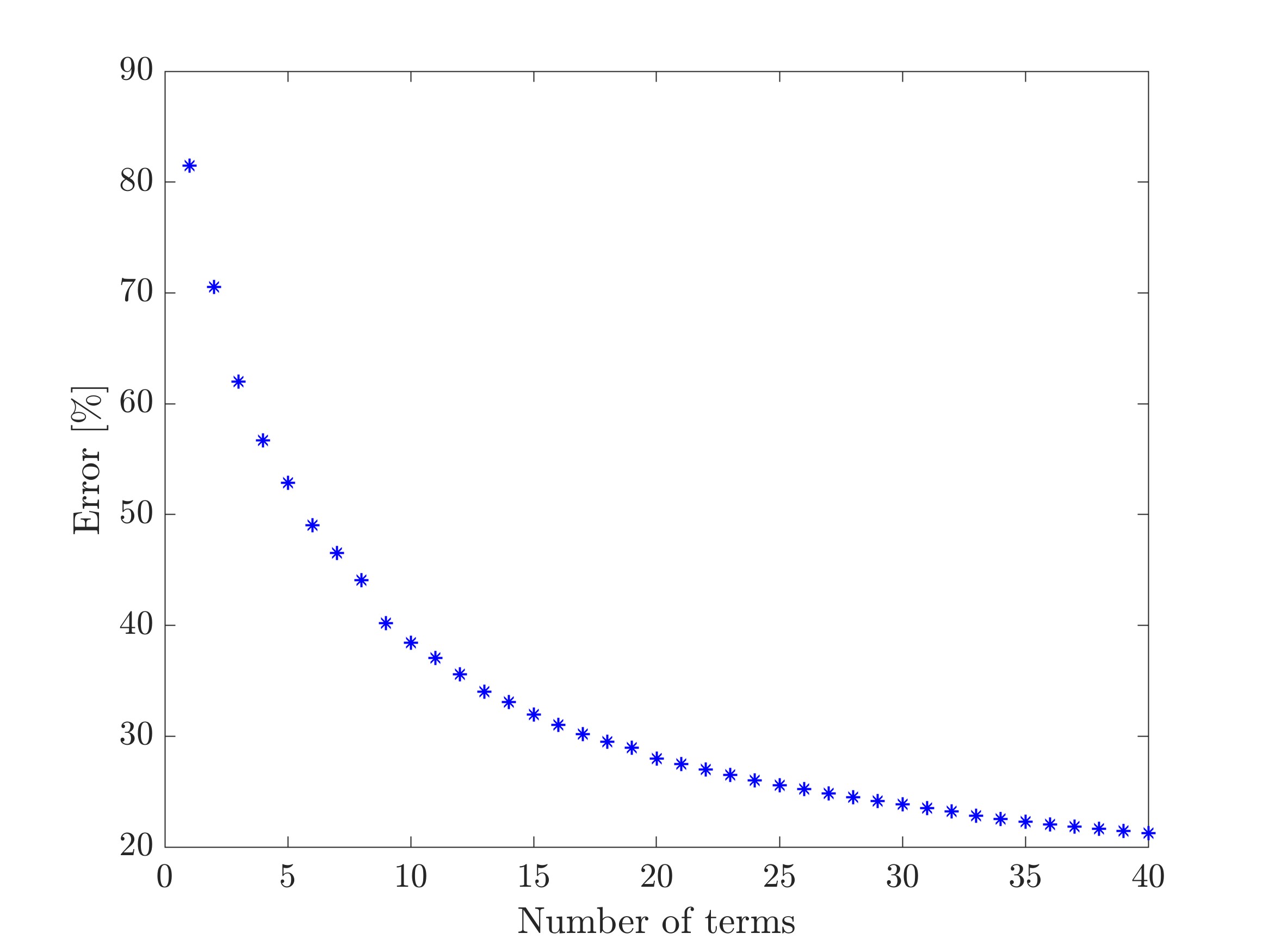}
\caption{Error versus number of submodes of the decomposition.}\label{fig:Error_complex}
\end{figure}

Here we considers the signal to be approximated as the one illustrated on Fig.~\ref{fig:Reference_sig_complex}. The input excitation considered is illustrated in Fig.~\ref{fig:Input_complex}. The input signal consists of a burst signal centered at $\omega = 0.62 \times 10^6 $ [rad/s].
After applying the signal approximation we obtain a reconstruction error of $20 \%$ by using $40$ submodes. The approximation compared with the reference signal is shown in Fig.~\ref{fig:comparison_complex}.
The error versus submodes is shown in Fig.~\ref{fig:Error_complex}.
%
%

\begin{figure}[H]
\centering
\includegraphics[width=0.7\textwidth]{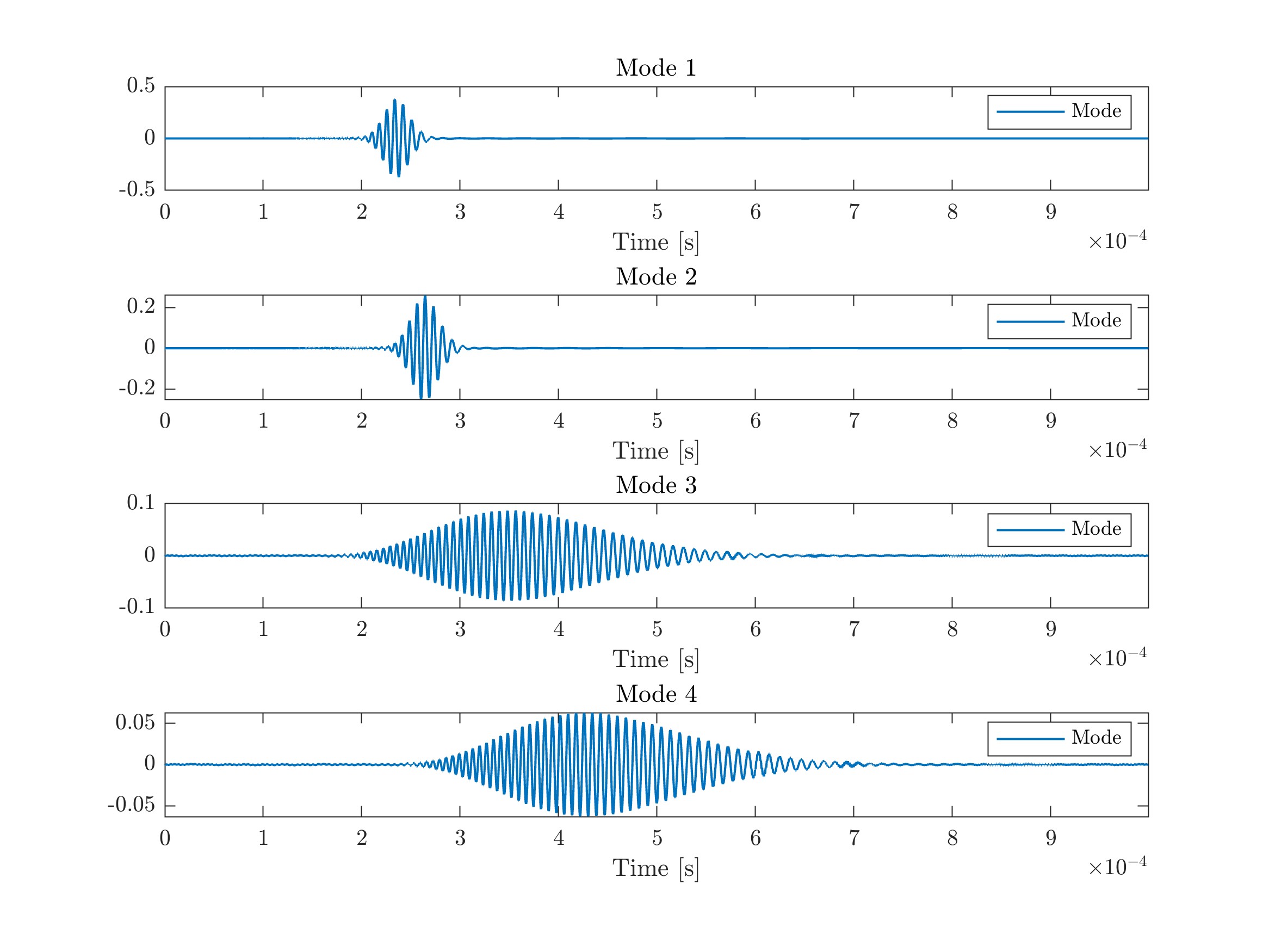}
\caption{Submodes of the decomposition.}\label{fig:submodes_complex}
\end{figure}

Finally, Fig.~\ref{fig:submodes_complex} shows the first $4$ submodes obtained through the proposed decomposition.

\subsection{Application to Thin-Plate Damage Localization}\label{sec:PISACMP_dam_det}

\begin{figure}[H]
\centering
\includegraphics[width=0.5\textwidth]{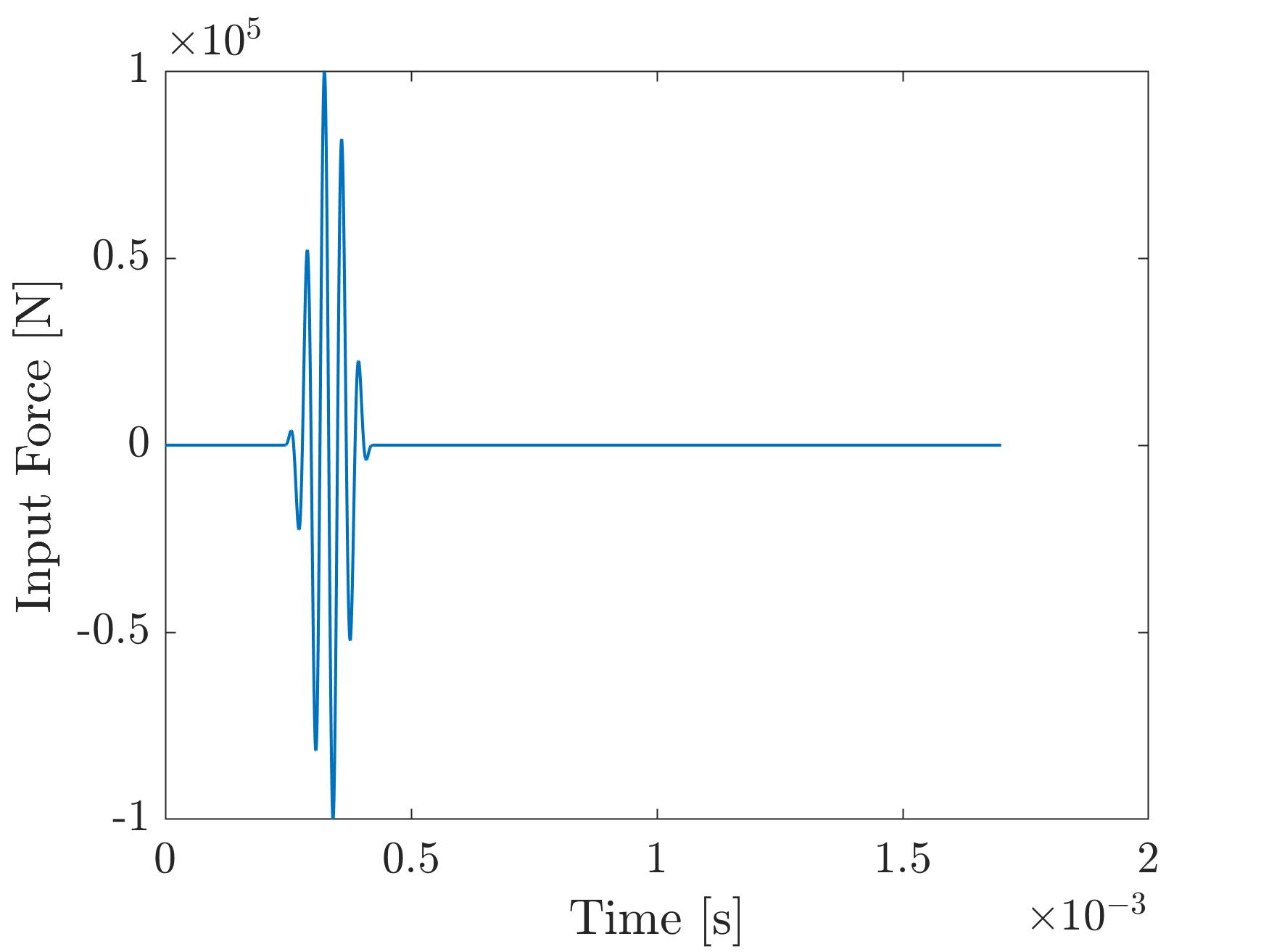}
\caption{Input signal (load) imposed on x and y directions at the center of the plate.}
\label{fig:Input_signal}
\end{figure}

\begin{figure}[H]
\centering
\includegraphics[width=1\textwidth]{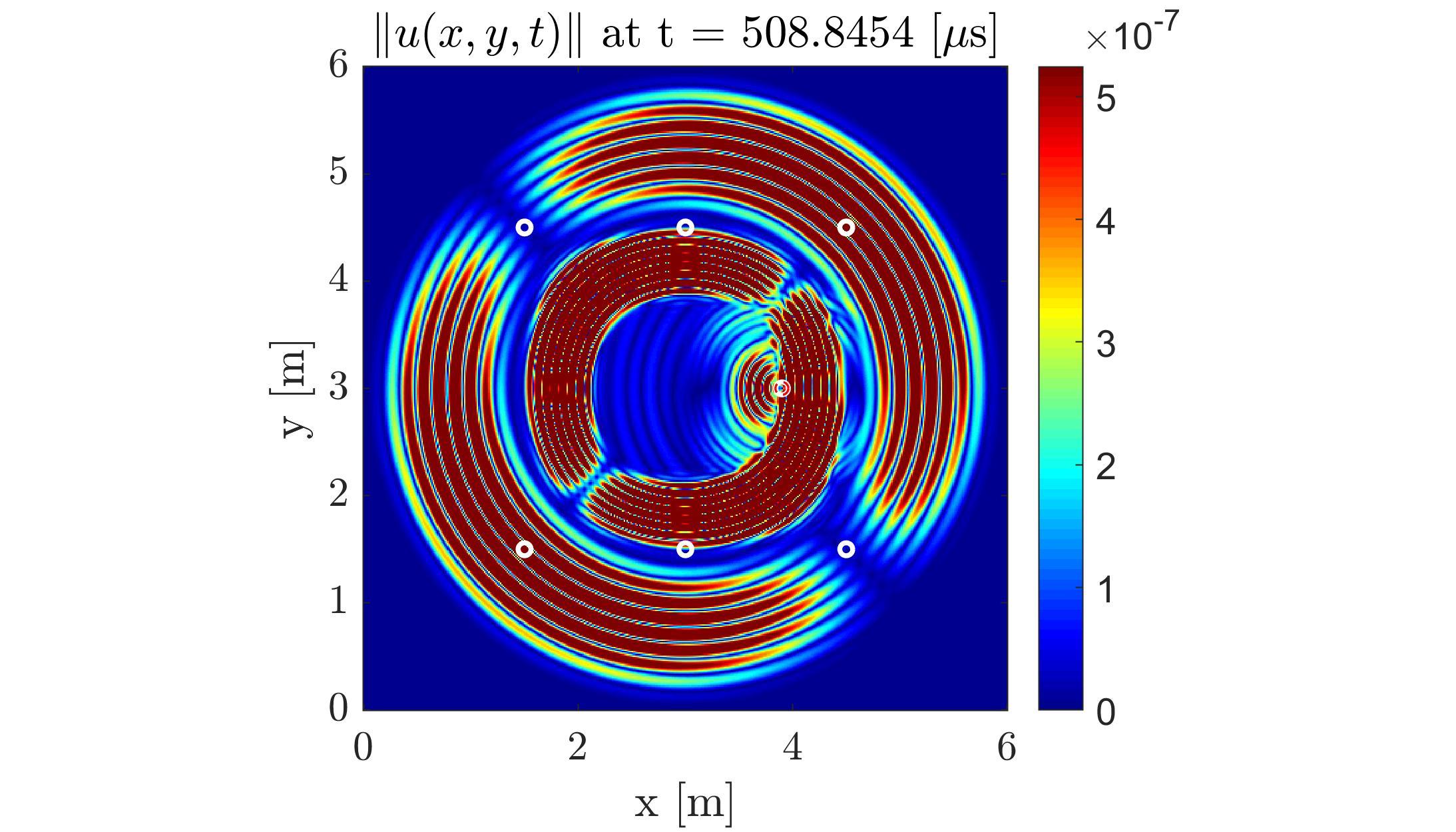}
\caption{Field of the displacement norm. Damage and sensors are depicted in red and white circles respectively.}
\label{fig:Field_damage}
\end{figure}

\begin{figure}[H]
\centering
\includegraphics[width=0.5\textwidth]{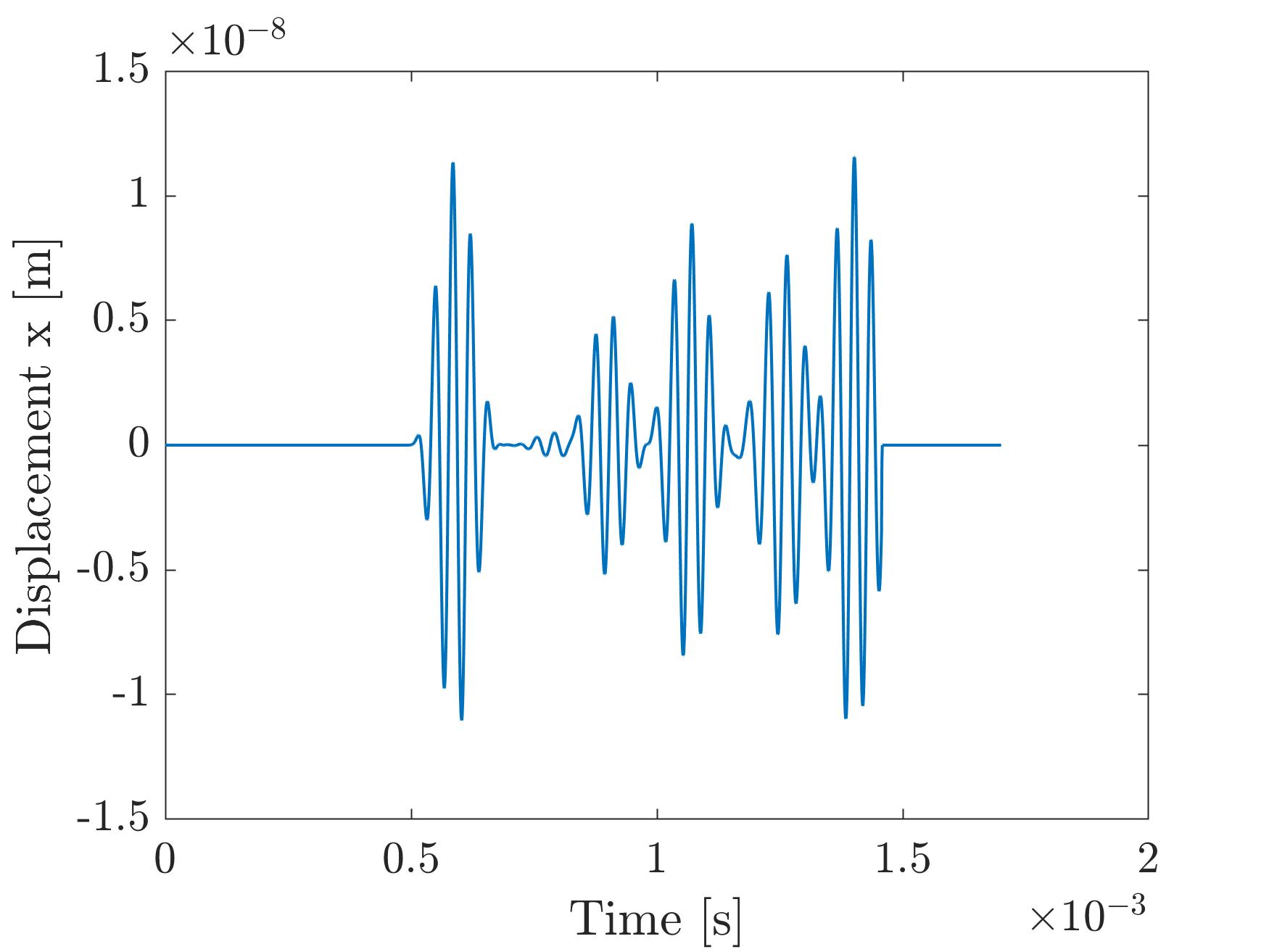}
\caption{Measured x-component displacement for the simulation of Fig.~\ref{fig:Field_damage}.}
\label{fig:Signal_u_x_damage}
\end{figure}

\begin{figure}[H]
    \centering
    \begin{subfigure}{0.48\textwidth}
        \centering
        \includegraphics[width=\linewidth]{undamaged_meassured.png}
        \caption{Measured signal.}
    \end{subfigure}
    \hfill
    \begin{subfigure}{0.48\textwidth}
        \centering
        \includegraphics[width=\linewidth]{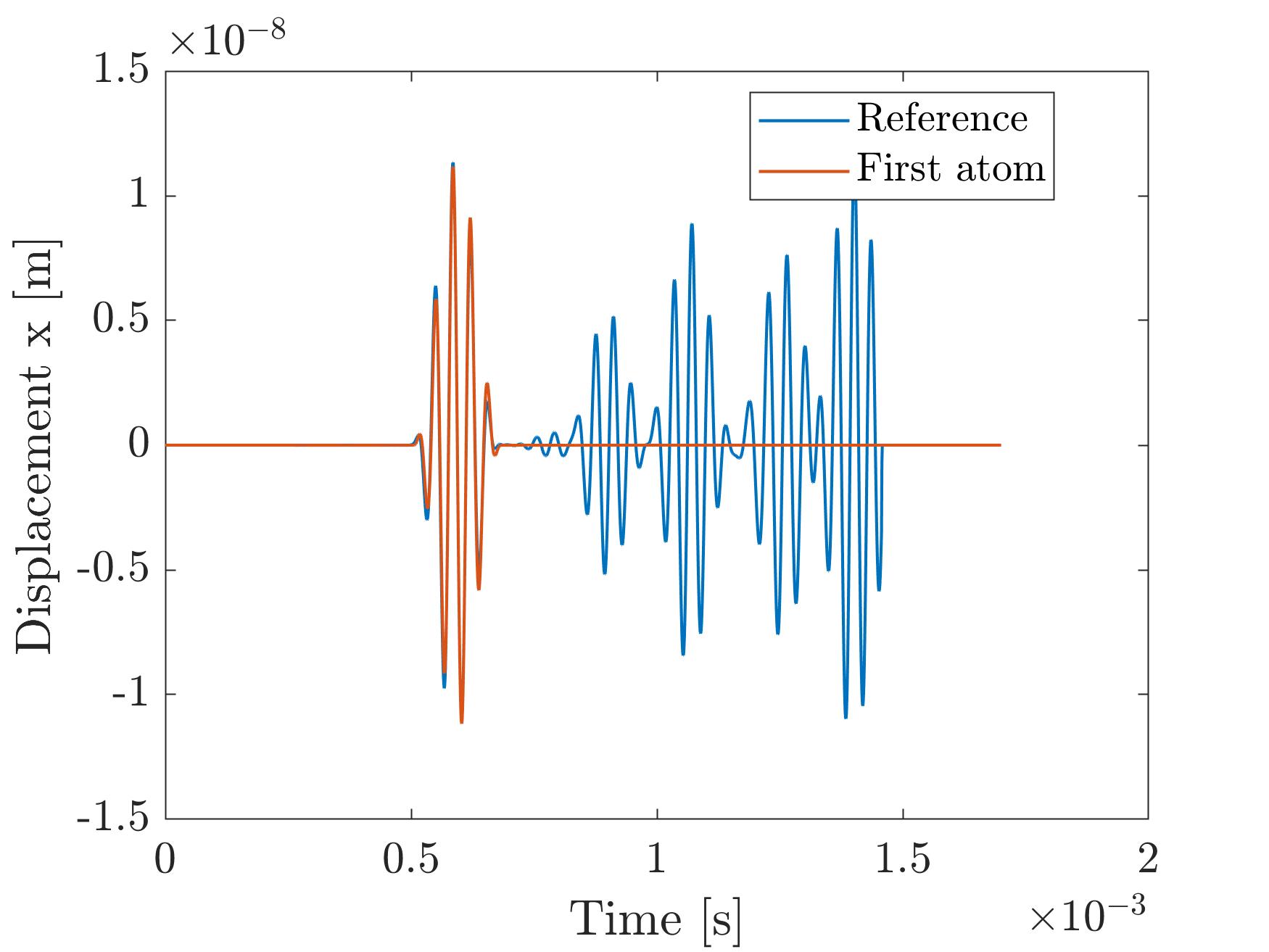}
        \caption{First arrival atom determination.}
    \end{subfigure}
    \caption{Measured signal at left and first arrival atom determination at right for the undamaged $s_i^{\text{undam}}(t)$ structure scenario.}
    \label{fig:first_arrival_undamaged_s3}
\end{figure}

Here, we consider a 2D square plate, with dimensions $L_x = L_y = 6$~[m]. The plate is considered to be isotropic with Young's and Poisson's modulus $E = 200$~[GPa] and $\nu = 0.3$ respectively. The plate is equipped with a single actuator located at its center, which generates a burst excitation signal centered at $20$~[kHz], consisting of five cycles as illustrated in Fig.~\ref{fig:Input_signal}, and $6$ sensors, $x_s = (L_x  \times 0.25,L_x \times 0.50, L_x \times 0.75)$ for $y_s = Ly \times 0.75$, and the other group of sensors at $x_s = (L_x  \times 0.25,L_x \times 0.50, L_x \times 0.75)$ for $y_s = L_y \times 0.25$. The whole process is simulated up to $T = 0.0017$~[s]. The damage is considered to be a square with its face equals $1$~cm.
\begin{figure}[H]
\centering
\includegraphics[width=0.4\textwidth]{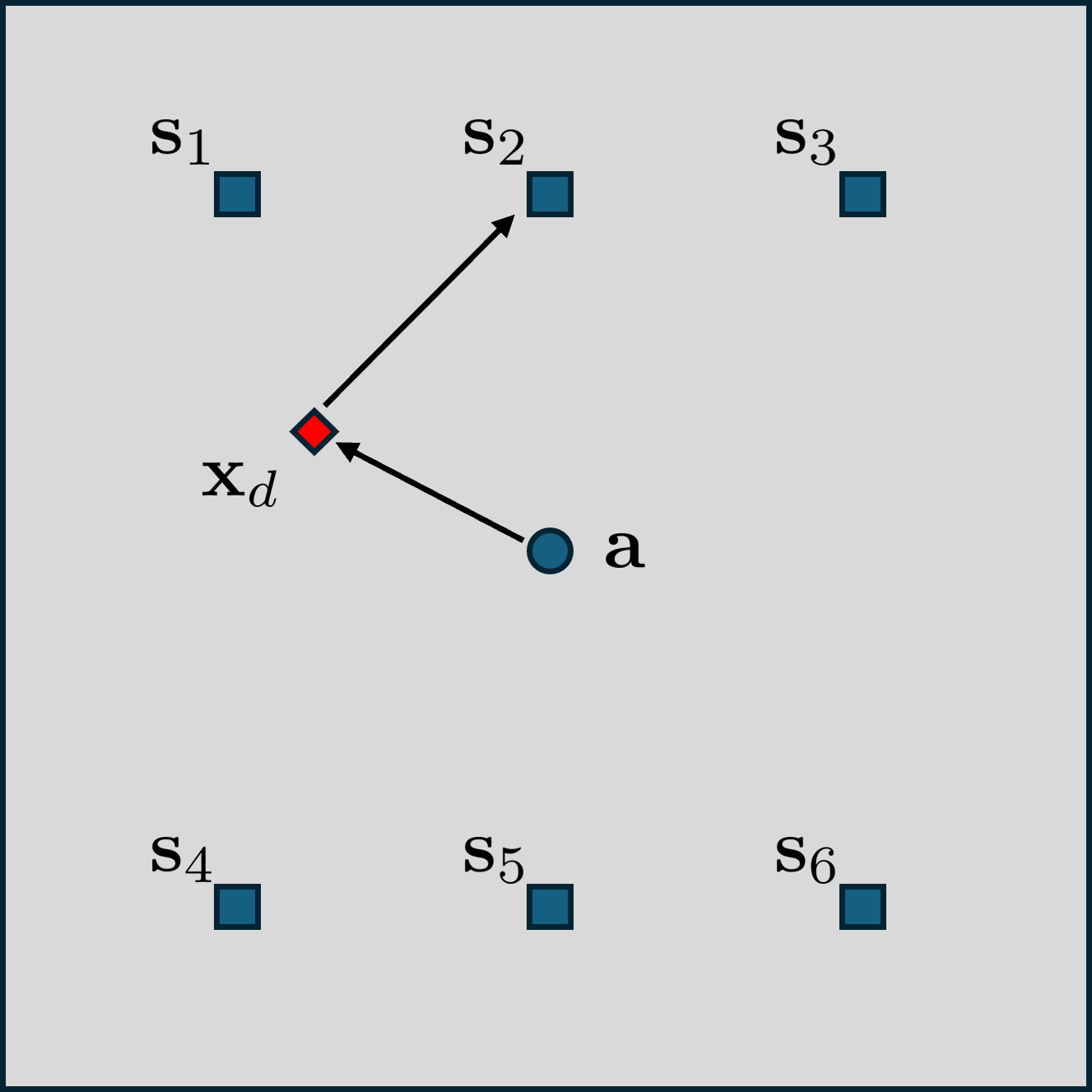}
\caption{Reference problem.}\label{fig:Ref_prob_num}
\end{figure}
Under this configuration, Fig.~\ref{fig:Field_damage} shows the field of the displacement norm at a specific instant. On the other hand, Fig.~\ref{fig:Signal_u_x_damage} shows the x-component of displacement measured at sensor $3$ location over the whole time domain.
By applying the PISACMP and determining the distance travelled following Section \ref{sec:dist_first_arrival}, the first atom obtained for sensor $3$ when measuring the undamaged structure is illustrated in Fig.~\ref{fig:first_arrival_undamaged_s3}.
%
%

\begin{figure}[h!]
\centering
\includegraphics[width=1\textwidth]{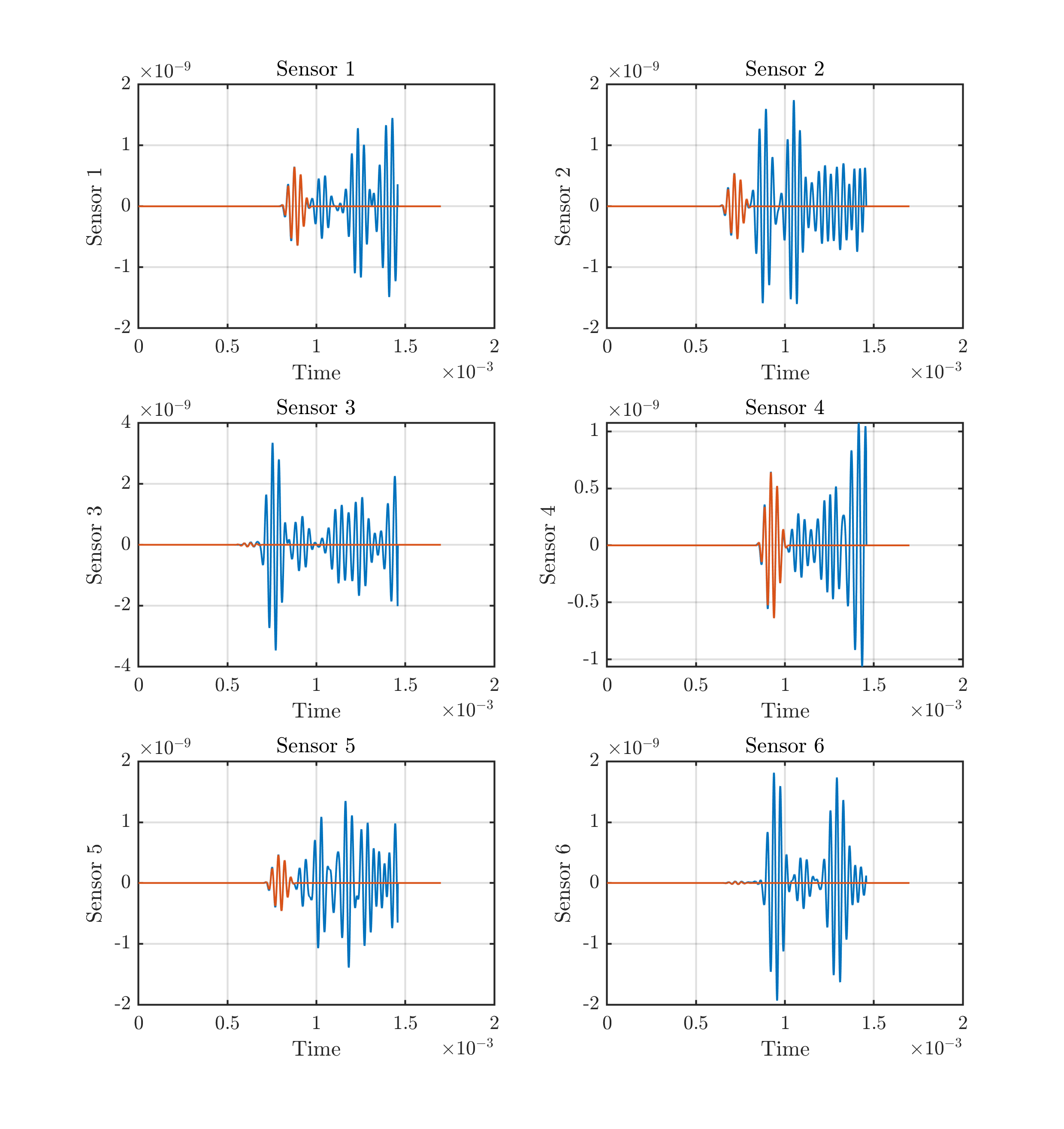}
\caption{Reference signal $s_i^{\text{dam}}(t) - s_i^{\text{undam}}(t)$ for $i=1,\cdots,6$ sensors in blue and first arrival atom determined in red.}
\label{fig:first_arrival_damaged_s3}
\end{figure}

By following the same procedure but now to the difference between the healthy and the signal measured with damage for the $6$ sensors, one obtain the results illustrated in Fig.~\ref{fig:first_arrival_damaged_s3}.

\begin{figure}[H]
    \centering
    \begin{subfigure}{0.48\textwidth}
        \centering
        \includegraphics[width=\linewidth]{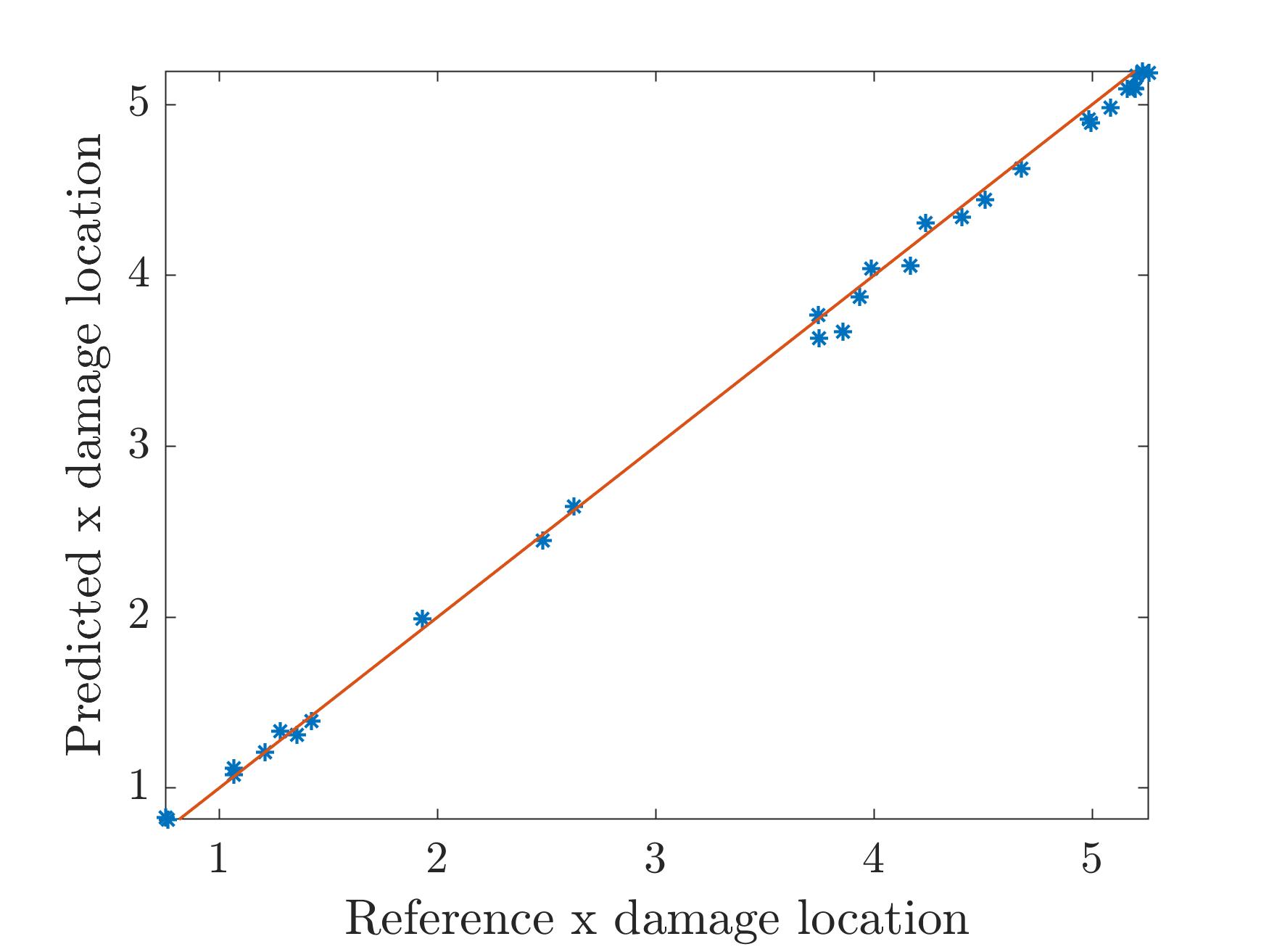}
        \caption{Reference vs predicted x-coordinate of damage.}
    \end{subfigure}
    \hfill
    \begin{subfigure}{0.48\textwidth}
        \centering
        \includegraphics[width=\linewidth]{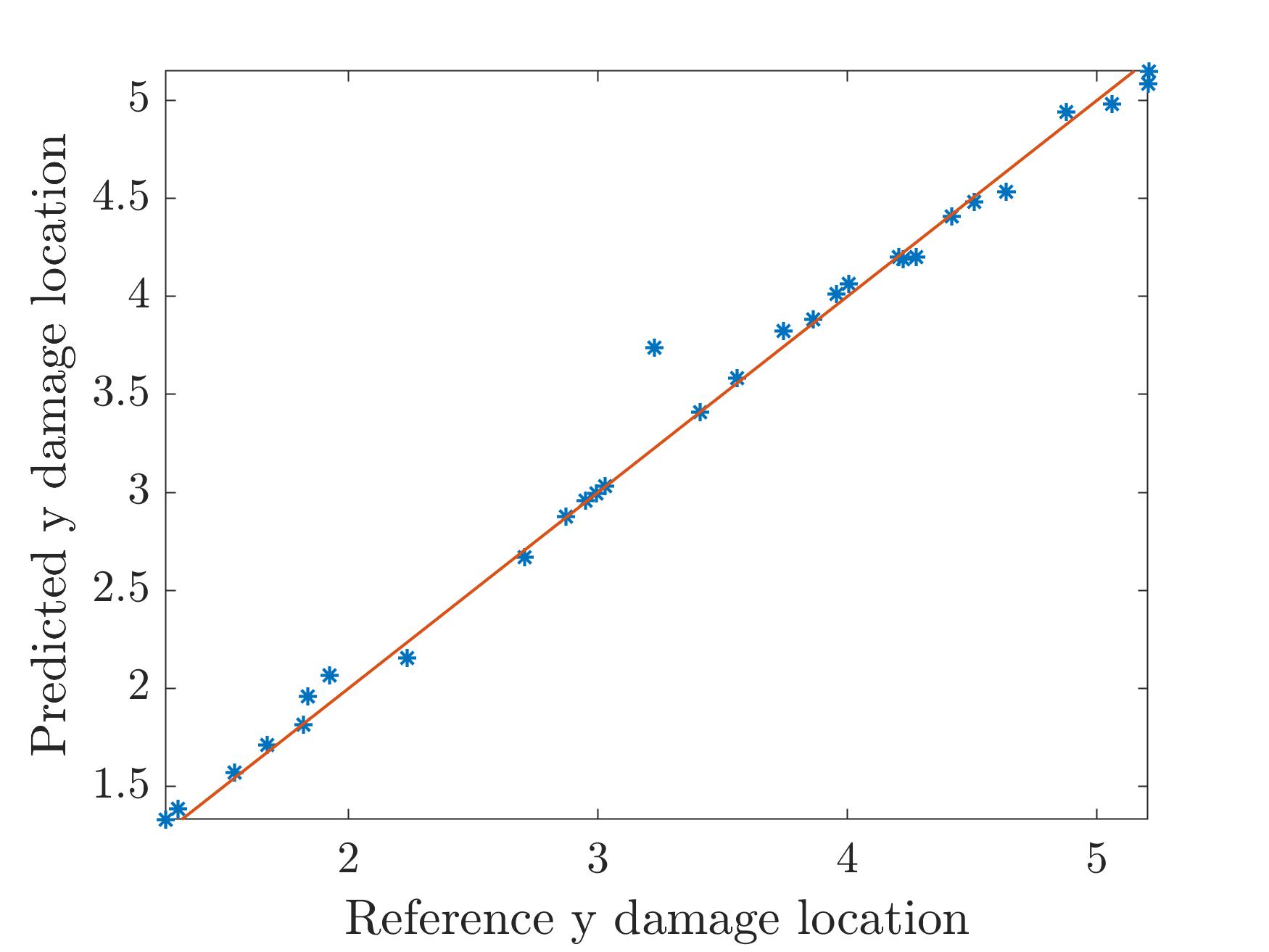}
        \caption{Reference vs predicted y-coordinate of damage.}
    \end{subfigure}
    \caption{PISACMP performance in damage location.}
    \label{fig:dam_location_x_y}
\end{figure}

Once the first arrival distances determined, for numerical simulations considering localized damage at different locations, the damage location method proposed in Section \ref{sec:dam_loc_framework} is applied. Figure \ref{fig:dam_location_x_y} illustrate the predicted $x$ and $y$ locations with respect to the reference locations when considering $30$ different simulations. The damage locations were selected using Latin Hypercube Sampling (LHS) of the x- and y-coordinate space.
The prediction results of damage location in terms of the averaged relative error defined below (where $x_d$ denote here any coordinate): 
\begin{equation}
\text{Error}
=
\frac{100}{N_s}
\sum_{i=1}^{N_s}
\left|
\frac{x^{\text{Ref}}_{d,i}-x^{\text{PISACMP}}_{d,i}}{x^{\text{Ref}}_{d,i}}
\right| \ [\%]
\end{equation}
with $N_s = 30$ different damage configurations considered, are summarized in Table~\ref{tab:PISACMP_loc_pred}.
\begin{table}[!ht]
\centering
\begin{tabular}{| c | c | }
\hline
Error $x$ coordinate $[\%]$ & Error $y$ coordinate $[\%]$ \\
\hline
2.35 & 2.26 \\ 
\hline
\end{tabular}
\caption{Averaged relative error for damage location prediction using PISACMP and Elliptical Localization approach.}
\label{tab:PISACMP_loc_pred}
\end{table}
As can be seen from Figure \ref{fig:dam_location_x_y} and Table \ref{tab:PISACMP_loc_pred}, the proposed algorithm, based purely on a Elliptical Localization approach using the distances obtained by the PISACMP method (Section \ref{sec:dam_loc_framework}), is able to correctly identify the damage location. This constitutes a major result of the present work. As a perspective for future research, this idea could be extended to curved geometries, where distances are no longer Euclidean and must instead be computed along geodesic paths on the structure.

\section{Conclusions and perspectives}\label{sec:concl_pers}

The new signal approximation technique proposed in this paper is based on the numerical determination of wavenumber functions in the frequency domain, amplitudes and distances traveling from the source signal to sensor location, which allows to extend the decomposition proposed in previous works, such as SAMPM and SACMPM \citep{rodriguez2025single}.

This decomposition has a strong physical meaning, since the extracted features correspond to the different distances traveled by the signals from the excitation source to the sensor, accounting for reflections at the boundaries of the domain as well as the presence of damage. Moreover, these results depend on the wavenumber functions numerically determined by the proposed algorithm. In addition, the proposed decomposition was used to perform damage detection through a geometrical Elliptical Localization approach, in which the first-arrival distances of the wave packets were employed. The x- and y-coordinates of the damage location were computed so as to best fit the distances associated with the scattered waves generated at the damage site.

As a perspective for future work, the proposed damage detection strategy should be applied and validated using real experimental data in order to assess its robustness and performance. In this context, the proposed methodology could be further enriched by combining simulated and experimental data within a Hybrid Twin framework \citep{chinesta2018virtual,moya2022digital,chinesta2023hybrid,rodriguez2023hybrid,ghnatios2024hybrid}. Such an approach could improve the accuracy of the estimated wavenumber functions and, consequently, the distances used to perform damage detection.
Furthermore, as an additional research perspective, the proposed strategy could be extended to curved geometries, where distances are no longer Euclidean and must instead be computed along geodesic paths on the structure. In this scenario, the PISACMP - Elliptical Localization approach can still be employed by considering geodesic distances, making the proposed PISACMP-based damage detection strategy a promising and interpretable methodology for addressing this challenging problem.




\bibliographystyle{elsarticle-harv}

\bibliography{references}


%
%
%

\end{document}